\documentclass[journal]{IEEEtran}
\usepackage{amsmath,amssymb}
\usepackage{graphicx}
\usepackage{booktabs}
\usepackage{tabularx}
\usepackage{tikz}
\usetikzlibrary{arrows.meta,positioning,fit,backgrounds,shapes.misc}
\usepackage{cite}
\usepackage{url}
\usepackage{array}
\usepackage{ragged2e}
\usepackage{microtype}

\newcolumntype{L}[1]{>{\RaggedRight\arraybackslash}p{#1}}

\begin{document}

\title{
Continuous Cuffless Blood Pressure Estimation Using PPG\\[-0.15em]
{\Large From Methods to Clinical Translation}
}

\author{Karthika~Shalini~Potail, Harshitha~Manjunatha, Hetvi~Kotadia, and Tooraj~Nikoubin\\
\textit{Department of Electrical and Computer Engineering}\\
\textit{The University of Texas at Dallas, Richardson, USA}\\
\{Karthika.Potail, Harshitha.Manjunatha, Hetvi.Kotadia, Tooraj.Nikoubin\}@UTDallas.edu}

\maketitle
\raggedbottom

\begin{abstract}
Continuous blood pressure (BP) monitoring could improve hypertension management by capturing changes that intermittent cuff measurements miss. Photoplethysmography (PPG) is widely explored for cuffless BP estimation because optical sensors are inexpensive, wearable, and already present in many consumer and clinical devices. PPG, however, does not measure arterial pressure: BP is inferred from waveform characteristics or pulse timing, which also depend on vascular tone, arterial stiffness, sensor contact, motion, and individual physiology. This survey reviews PPG-based cuffless BP estimation along the full measurement chain, covering physiological mechanisms, waveform features, pulse arrival time (PAT), pulse transit time (PTT), machine-learning methods, datasets, calibration, validation, deployment, bias, and clinical translation. Studies are assessed by the strength of evidence they provide rather than by reported error alone, with particular attention to subject and temporal leakage, dataset imbalance, calibration dependence and drift, motion artifacts, and the gap between benchmark accuracy and reliable real-world measurement. The review finds that progress depends on more than better prediction algorithms. Future work should prioritize physiologically informed models, diverse longitudinal datasets, transparent reporting of calibration requirements and failure rates, stronger evaluation protocols, and studies that demonstrate practical value in clinical care.
\end{abstract}

\section{Introduction}
\label{sec:intro}

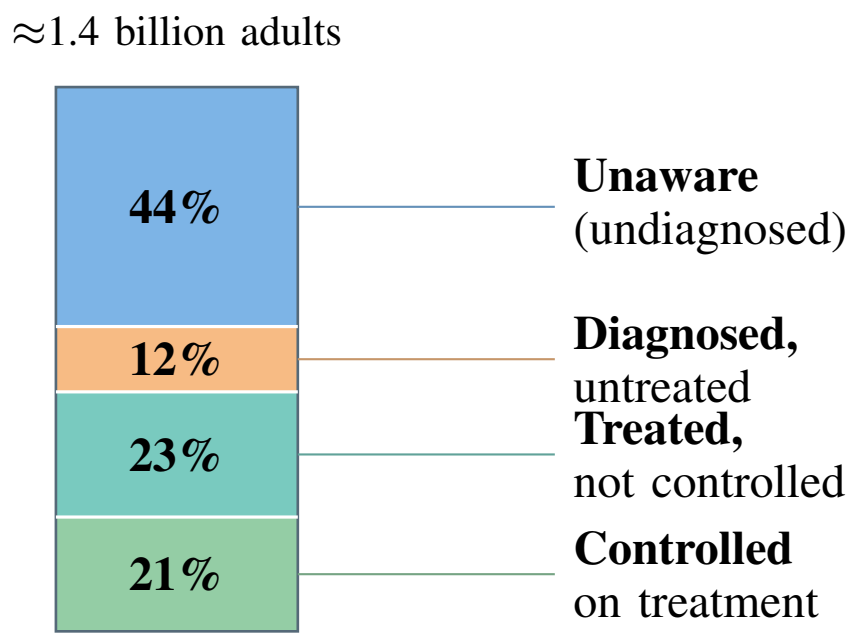
\begin{figure}[t]

\centering

\begin{tikzpicture}[
    every node/.style={font=\small},
]

\def\barheight{3.6}
\def\barwidth{1.6}

\definecolor{unaware}{RGB}{125,180,230}
\definecolor{untreated}{RGB}{247,187,132}
\definecolor{notcontrolled}{RGB}{121,202,191}
\definecolor{controlled}{RGB}{148,205,165}
\definecolor{bordercolor}{RGB}{85,105,120}

\fill[controlled]
    (0,0) rectangle
    (\barwidth,0.21*\barheight);

\node[
    font=\small\bfseries,
    black
] at (\barwidth/2,0.105*\barheight)
{21\%};

\fill[notcontrolled]
    (0,0.21*\barheight) rectangle
    (\barwidth,0.44*\barheight);

\node[
    font=\small\bfseries,
    black
] at (\barwidth/2,0.325*\barheight)
{23\%};

\fill[untreated]
    (0,0.44*\barheight) rectangle
    (\barwidth,0.56*\barheight);

\node[
    font=\small\bfseries,
    black
] at (\barwidth/2,0.50*\barheight)
{12\%};

\fill[unaware]
    (0,0.56*\barheight) rectangle
    (\barwidth,\barheight);

\node[
    font=\small\bfseries,
    black
] at (\barwidth/2,0.78*\barheight)
{44\%};

\draw[
    bordercolor,
    line width=0.5pt
]
    (0,0) rectangle
    (\barwidth,\barheight);

\draw[white, line width=0.6pt]
    (0,0.21*\barheight) --
    (\barwidth,0.21*\barheight);

\draw[white, line width=0.6pt]
    (0,0.44*\barheight) --
    (\barwidth,0.44*\barheight);

\draw[white, line width=0.6pt]
    (0,0.56*\barheight) --
    (\barwidth,0.56*\barheight);

\node[
    right,
    align=left,
    font=\small,
    anchor=west
] (l1) at
(\barwidth+1.7,0.105*\barheight)
{\textbf{Controlled}\\on treatment};

\node[
    right,
    align=left,
    font=\small,
    anchor=west
] (l2) at
(\barwidth+1.7,0.325*\barheight)
{\textbf{Treated,}\\not controlled};

\node[
    right,
    align=left,
    font=\small,
    anchor=west
] (l3) at
(\barwidth+1.7,0.50*\barheight)
{\textbf{Diagnosed,}\\untreated};

\node[
    right,
    align=left,
    font=\small,
    anchor=west
] (l4) at
(\barwidth+1.7,0.78*\barheight)
{\textbf{Unaware}\\(undiagnosed)};

\draw[controlled!80!black, thin]
    (\barwidth,0.105*\barheight) --
    (l1.west);

\draw[notcontrolled!80!black, thin]
    (\barwidth,0.325*\barheight) --
    (l2.west);

\draw[untreated!80!black, thin]
    (\barwidth,0.50*\barheight) --
    (l3.west);

\draw[unaware!80!black, thin]
    (\barwidth,0.78*\barheight) --
    (l4.west);

\node[
    above,
    font=\footnotesize
] at
(\barwidth/2,\barheight+0.15)
{$\approx$1.4 billion adults};

\end{tikzpicture}

\caption{Approximate global hypertension status among adults aged 30--79, based on WHO estimates \cite{who2024}.}
\label{fig:burden}

\end{figure}

Blood pressure (BP) is central to cardiovascular risk assessment and to the diagnosis and treatment of hypertension \cite{kannel1996,whelton2018}. The World Health Organization estimates that 1.4 billion adults aged 30--79 had hypertension in 2024. Only about one in five of them, some 320 million people, had their BP under control, and close to half did not know they had the condition \cite{who2024}. Fig.~\ref{fig:burden} shows how this population divides between those who are unaware, diagnosed but untreated, treated without reaching control, and controlled. Most of the shortfall lies in detection and long-term management, not in the availability of effective drugs.

\begin{figure}[!t]
\hspace*{-0.55cm}
\includegraphics[
    width=1.03\columnwidth
]{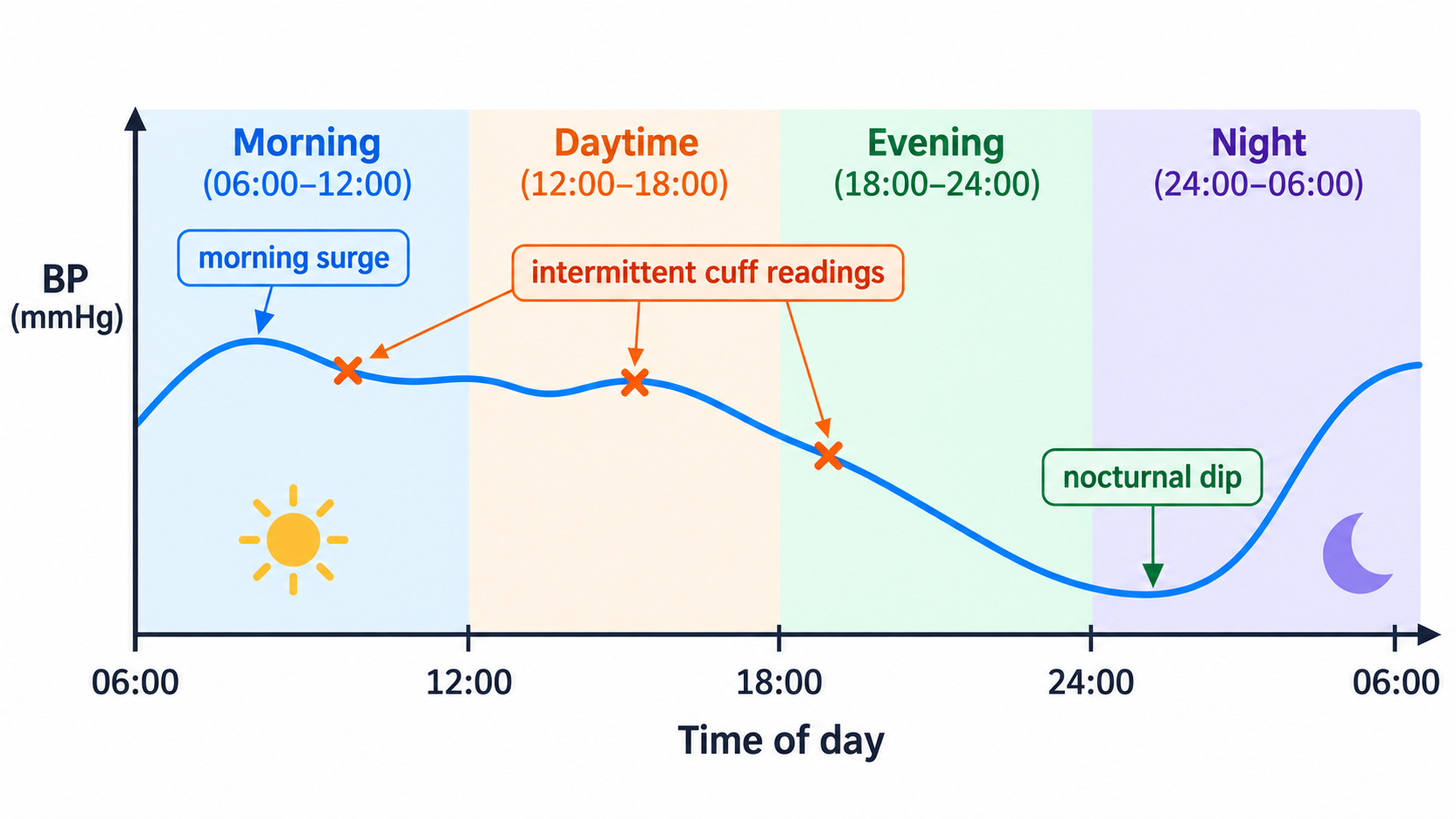}
\caption{24-hour BP pattern showing nocturnal dipping and a morning rise \cite{obrien2013}.}
\label{fig:diurnal}
\end{figure}

The cuff remains the clinical reference. It is supported by outcome evidence, measurement guidelines, and established validation protocols \cite{muntner2019,pickering2005,obrien2013,stergiou2018_universal,iso81060}, but it samples BP only intermittently. Office and home readings capture a few moments of the day, and ambulatory monitoring, although denser, still depends on repeated inflation. That matters because BP follows a daily rhythm. In many people it falls by approximately 10--20\% during sleep and rises again in the early morning, and non-dipping, reverse dipping, and an exaggerated morning rise have all been associated with increased cardiovascular risk \cite{obrien2013}.Fig.~\ref{fig:diurnal} illustrates this limitation: the three cuff readings do not capture the morning peak or nocturnal trough.

Cuffless monitoring aims to fill these gaps with sensors that can be worn continuously. PPG is a widely investigated candidate because optical sensors are inexpensive and already built into smartwatches, rings, smartphones, and clinical pulse oximeters \cite{tamura2025,zhao2023,allen2007,sun2016,charlton2022}.

PPG does not measure arterial pressure directly. It records blood-volume changes in peripheral tissue, and BP has to be inferred from the shape of that signal or from pulse arrival time (PAT) and pulse transit time (PTT). All of these also depend on arterial stiffness, vascular tone, autonomic activity, and individual physiology \cite{mukkamala2015,payne2006}, and the waveform itself changes with motion, temperature, sensor contact, and perfusion. The mapping from PPG to BP differs from person to person and drifts within a person over time.

Machine-learning and deep-learning models have reported low errors on public benchmarks, but benchmark performance alone does not establish generalizability. Accuracy depends on the population and BP range of the dataset, on signal quality, on how training and test data are separated, and on whether the model is calibrated to each user \cite{sharman2020,wang2023pulsedb}. Reported accuracy can be inflated when recordings from the same subject appear in both training and test sets, or because it is recalibrated against a cuff so often that it adds little to the cuff. Few commercial devices have been validated at all \cite{cox2022}, and expert groups have argued that the standards used for cuff devices are not adequate for cuffless ones \cite{mukkamala2021}.

Several reviews already cover this field. El-Hajj and Kyriacou \cite{elhajj2020} and Qin et al. \cite{qin2022} catalogue machine-learning architectures by model family, Zhao et al. \cite{zhao2023} survey sensing hardware, Charlton et al. \cite{charlton2022} review wearable PPG for cardiovascular monitoring more broadly, and Islam et al. \cite{islam2022} pool validation results for commercial wearables. This survey differs by evaluating studies according to the strength of evidence supporting their claims. Cuffless BP estimation is evaluated as a measurement problem spanning physiology, signal processing, modeling, validation, and clinical translation.

The paper makes four contributions. It classifies studies by the strength of their evidence, from physiological plausibility to clinical utility. It interprets reported accuracy in light of subject independence, temporal separation, calibration, and BP-range coverage. It brings together the growing evidence on cross-dataset generalization, ambulatory performance, and commercial-device validation, which shows how far benchmark accuracy overstates real-world accuracy. Finally, it assesses newer directions, including physics-informed models, on-device inference, remote PPG, and federated and self-supervised learning, against the same criteria.

Section~\ref{sec:pipeline} outlines the estimation pipeline, and Sections~\ref{sec:physio}--\ref{sec:taxonomy} cover its physiological basis, waveform features, and estimation methods. Sections~\ref{sec:mlpitfalls}--\ref{sec:wearable} examine how these methods are evaluated and where they fail. Section~\ref{sec:clinical} turns to clinical translation, Section~\ref{sec:future} sets out research priorities, Section~\ref{sec:limitations} states the limitations of this review, and Section~\ref{sec:conclusion} concludes.

\section{Overview of the PPG-Based Cuffless BP Pipeline}
\label{sec:pipeline}
\begin{figure}[t]
\centering
\includegraphics[
    width=0.98\columnwidth,
    trim=0.25cm 0.25cm 0.25cm 0.25cm,
    clip
]{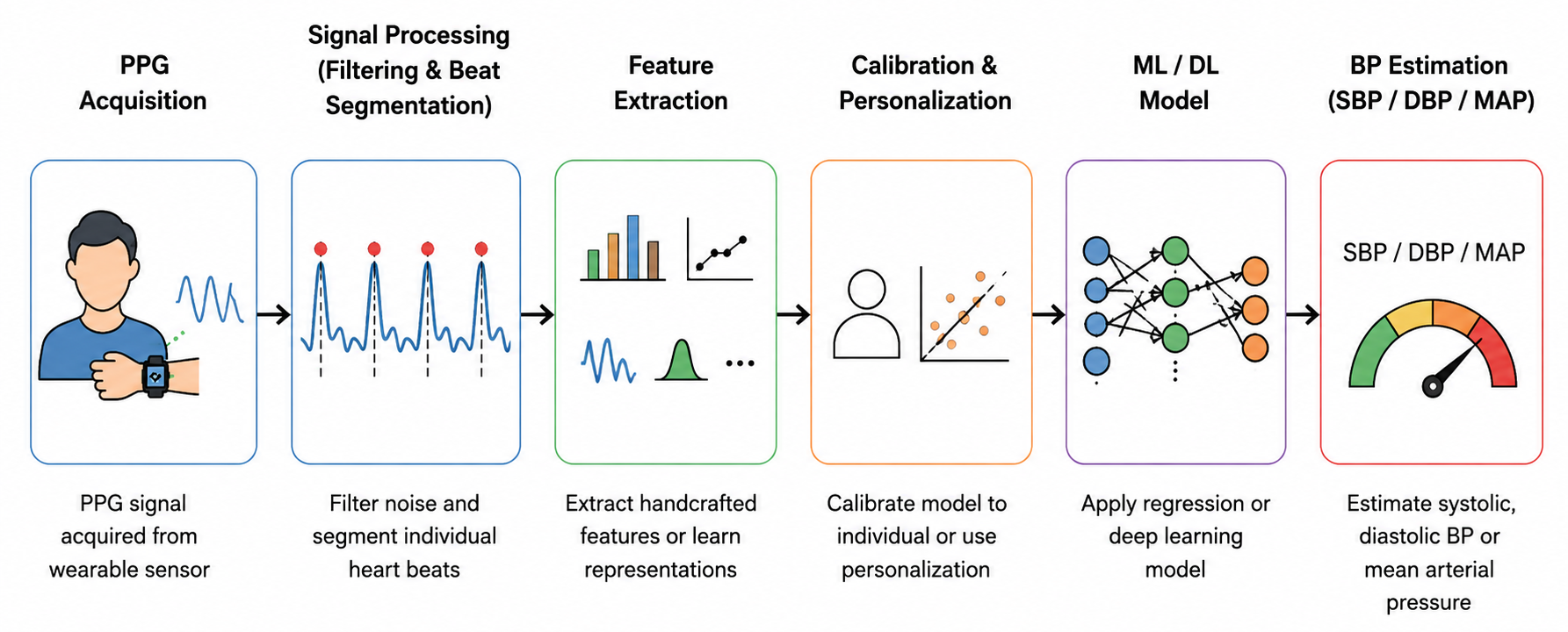}
\caption{Overview of a modern PPG-based cuffless blood pressure estimation pipeline.}
\label{fig:pipeline}
\end{figure}

Most PPG-based systems share the processing chain in Fig.~\ref{fig:pipeline} \cite{zhao2023,tamura2025,qin2022}. A wearable sensor records the PPG, which is filtered and cut into individual beats, with poor-quality segments discarded. Features are then extracted from each beat by hand or learned by a neural network. Because the PPG--BP relationship differs between people, the mapping is usually calibrated or personalized, and a regression or deep-learning model finally outputs SBP, DBP, or MAP.

Errors introduced during acquisition or preprocessing can propagate through subsequent stages of the pipeline.  Poor contact corrupts acquisition, filtering and segmentation can move the landmarks that features depend on, and model generalization is limited by the diversity of the training data and the calibration conditions used during development. The sections that follow take the stages in order, starting with the physiology that connects an optical signal to arterial pressure.

\section{Physiological Mechanisms of Cuffless BP Estimation}
\label{sec:physio}
\subsection{PPG as an Optical Surrogate}
\begin{figure}[t]
\centering
\includegraphics[width=\linewidth]{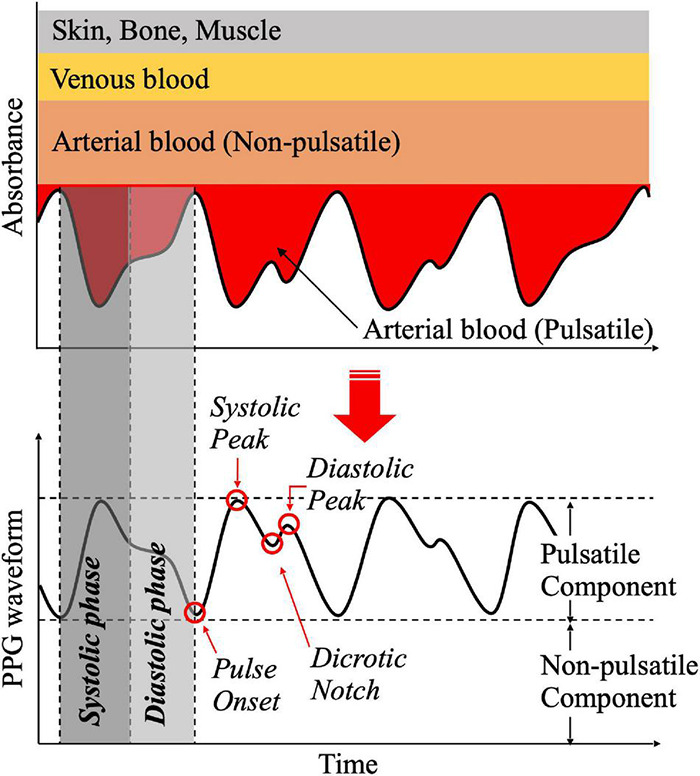}
\caption{PPG signal components and characteristic waveform morphology. Reproduced from \cite{park2022}.}
\label{fig:ppgmorph}
\end{figure}

PPG measures how much light tissue absorbs or reflects as its blood volume changes with each heartbeat \cite{allen2007,nitzan2014,sun2016}. The upper panel of Fig.~\ref{fig:ppgmorph} shows where that light goes. Skin, bone, muscle, venous blood, and non-pulsatile arterial blood absorb a roughly constant amount; only the pulsatile arterial component varies from beat to beat, and it is typically a small fraction of the total. The signal is therefore modeled as
\begin{equation}
PPG(t) = DC + AC(t)
\label{eq:ppg}
\end{equation}
where $DC$ is the baseline (non-pulsatile) component and $AC(t)$ is the pulsatile arterial component.

The lower panel of Fig.~\ref{fig:ppgmorph} shows the resulting waveform, divided into a systolic phase, when the vascular bed fills, and a diastolic phase, when it empties. Each pulse carries four landmarks: the pulse onset at the start of the upstroke, the systolic peak at maximum volume, the dicrotic notch associated with aortic valve closure, and the diastolic peak associated with waves reflected from the periphery. The notch is often weak or missing at the wrist. Pulse shape reflects cardiac ejection, vascular compliance, peripheral resistance, and wave reflection, but also sensor optics and local tissue properties \cite{elgendi2012,park2022,lin2020}. Because the pulsatile part rides on a much larger baseline, anything that shifts the baseline, such as a change in perfusion or contact pressure, also changes the apparent pulse amplitude. Vasoconstriction, vasodilation, temperature, sympathetic activation, local compression, sensor displacement, and processing artifacts can all alter the waveform without any change in BP. PPG is best understood as an indirect surrogate: it carries information about BP, mixed with several confounders.

\subsection{PAT, PTT, and PEP}
\begin{figure}[t]
\centering

\resizebox{0.92\columnwidth}{!}{%
\begin{tikzpicture}[
    every node/.style={font=\footnotesize},
    wave/.style={line width=0.9pt},
    guide/.style={dashed, thin},
    measure/.style={<->, >=Latex, thin}
]

\def\xstart{0}
\def\xend{5.7}

\node[anchor=east] at (-0.25,4.4) {ECG};

\draw[wave]
    (\xstart,4.4) -- (0.75,4.4)
    -- (0.90,5.05)
    -- (1.05,4.05)
    -- (1.28,4.4)
    -- (\xend,4.4);

\node[anchor=east] at (-0.25,2.75) {proximal PPG};

\draw[wave]
    (\xstart,2.55)
    .. controls (0.90,2.55) and (1.35,2.65) .. (1.75,3.05)
    .. controls (2.00,3.35) and (2.18,3.35) .. (2.38,3.10)
    .. controls (2.75,2.70) and (3.50,2.58) .. (\xend,2.55);

\node[anchor=east] at (-0.25,1.15) {distal PPG};

\draw[wave]
    (\xstart,0.95)
    .. controls (1.10,0.95) and (1.85,1.05) .. (2.40,1.45)
    .. controls (2.85,1.82) and (3.15,2.05) .. (3.45,2.05)
    .. controls (3.80,2.05) and (4.15,1.50) .. (4.55,1.10)
    .. controls (4.85,0.95) and (5.20,0.95) .. (\xend,0.95);

\draw[guide] (0.90,4.95) -- (0.90,0.55);

\draw[guide] (1.85,4.75) -- (1.85,0.55);

\draw[guide] (2.10,3.35) -- (2.10,0.55);

\draw[guide] (3.45,2.10) -- (3.45,0.55);

\draw[measure]
    (0.90,5.20) -- node[above=1pt]{PEP} (1.85,5.20);

\draw[measure]
    (2.10,0.40) -- node[below=1pt]{PTT} (3.45,0.40);

\draw[measure]
    (0.90,-0.05) -- node[below=1pt]{PAT} (3.45,-0.05);

\end{tikzpicture}%
}

\caption{Conceptual distinction between PEP, PTT, and PAT. PAT includes
the electromechanical delay represented by PEP in addition to vascular
transit time, making it less specific to BP than idealized PTT.}

\label{fig:peppat}
\end{figure}
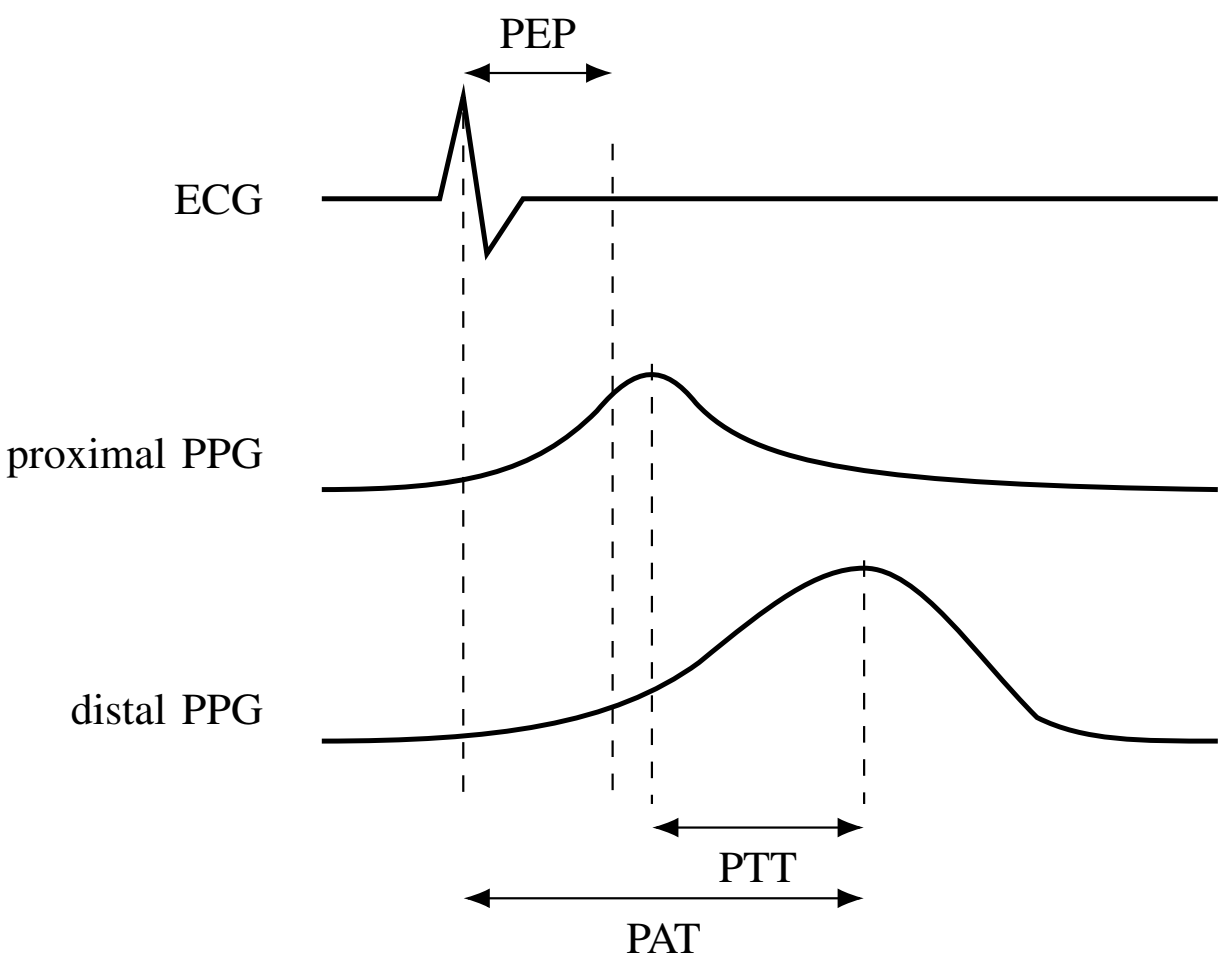

Pulse-timing methods use the time the pulse wave needs to reach a peripheral site. Fig.~\ref{fig:peppat} separates the three intervals involved. The pre-ejection period (PEP) is the electromechanical delay between electrical activation, marked by the ECG, and the opening of the aortic valve. PTT is the time the pulse takes to travel between a proximal and a distal arterial site. PAT, usually measured from the ECG R-peak to a landmark on the distal PPG, spans both:
\begin{equation}
PAT = PEP + PTT
\label{eq:pat}
\end{equation}
Measuring PTT directly requires two synchronized sensors at different arterial sites, which is difficult in a single wearable. PAT needs only an ECG and one PPG, but it is less specific to BP: PEP changes with contractility, stress, autonomic activity, and medication independently of vascular transit \cite{mukkamala2015,payne2006}, so a change in PAT can originate in the heart instead of the arteries.

\subsection{Pulse Wave Velocity and Wave Reflection}
\begin{figure}[t]
\centering
\includegraphics[width=\linewidth]{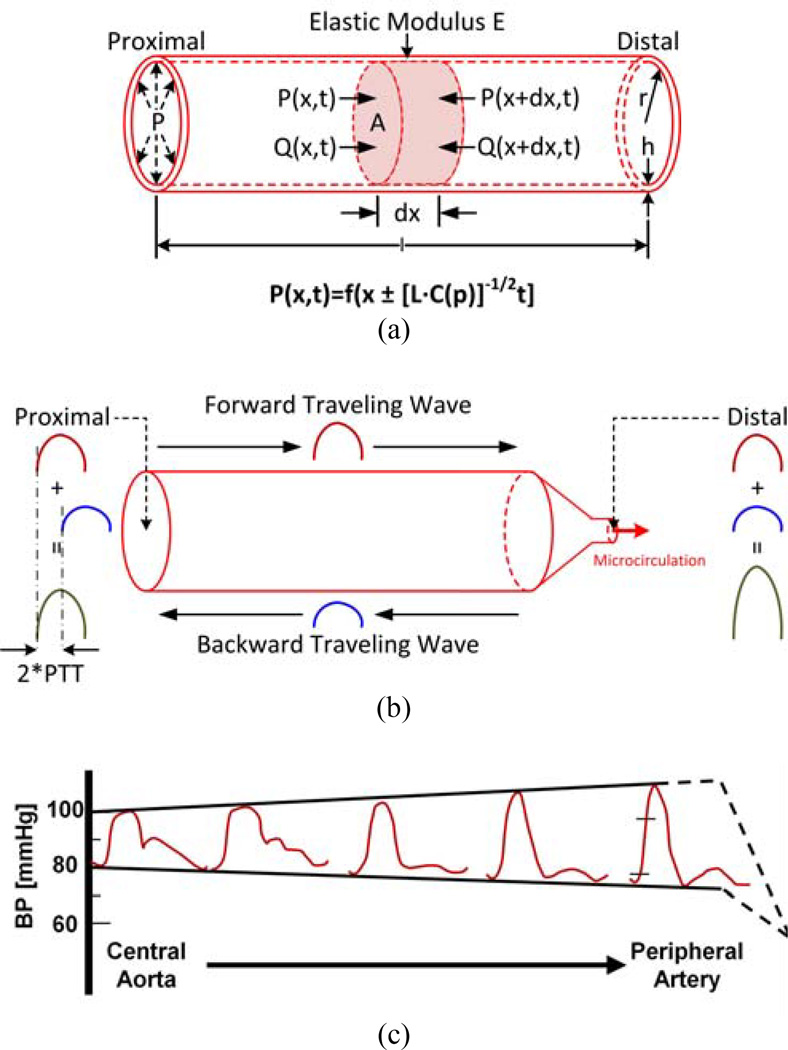}
\caption{Pulse wave propagation, wave reflection, and pressure amplification in the arterial tree. Adapted from \cite{mukkamala2015}.}
\label{fig:wavepropagation}
\end{figure}

PTT-based methods \cite{poon2006,chen2009,mukkamala2015,ding2017} rely on the dependence of pulse wave velocity (PWV) on the elasticity of the arterial wall. Panel (a) of Fig.~\ref{fig:wavepropagation} models the artery as an elastic tube, for which the Moens--Korteweg equation gives
\begin{equation}
PWV = \sqrt{\frac{Eh}{2r\rho}}
\label{eq:pwv}
\end{equation}
where $E$ is the arterial elastic modulus, $h$ the vessel wall thickness, $r$ the arterial radius, and $\rho$ the blood density. The same panel writes propagation in transmission-line form, with pressure traveling as $P(x,t)=f(x \pm [L\,C(p)]^{-1/2}t)$, so that
\begin{equation}
PWV = \frac{1}{\sqrt{L\,C(p)}}
\label{eq:pwvlc}
\end{equation}
where $L$ is the blood inertance and $C(p)$ the pressure-dependent arterial compliance per unit length \cite{mukkamala2015}. Compliance falls as pressure rises, so higher BP gives faster propagation and a shorter PTT. That pressure dependence is what makes PTT informative. Its variation from one vessel and one person to another is what makes the relationship subject-specific, and arterial stiffness, vascular tone, aging, medication, disease, posture, and exercise all shift PWV independently of acute BP \cite{payne2006,gesche2012,ding2019}.

Panels (b) and (c) add two complications. In panel (b), the waveform at any point is the sum of a forward wave from ventricular ejection and a backward wave reflected from the microcirculation. Near the proximal end the two are separated by roughly twice the transit time; near the distal end they overlap. The same heartbeat thus produces differently shaped pulses at different sites, and reflection alters pulse shape and timing independently of absolute BP. Panel (c) shows that as the pulse travels from the central aorta to a peripheral artery, the systolic peak rises while diastolic pressure falls slightly, widening the pulse pressure, so a peripheral measurement is not a direct reading of central pressure. No single timing-to-BP conversion holds across individuals, which is why PTT-based estimation almost always needs calibration or personalization.

\section{PPG Signal Morphology and Feature Representation}
\label{sec:morphology}
Feature-based methods summarize each pulse with a handful of measurements, such as systolic amplitude, upstroke slope, rise time, pulse width, pulse area, derivative peaks, augmentation-related indices, and dicrotic-notch timing, and feed them to a regression model \cite{allen2007,elgendi2012,lin2020}. The features are cheap to compute and easy to interpret, which suits wearable hardware. Each of them, however, depends on a landmark that something other than BP can move.

The features fall into four families, each tied to part of the physiology above. Amplitude and area features describe how much the vascular bed fills and so reflect perfusion and compliance. Timing features, including rise time, pulse width, and notch timing, describe how quickly it fills and empties and are linked to propagation. Shape features such as augmentation indices capture the contribution of reflected waves. Derivative features, taken from the first and second derivatives of the PPG, sharpen inflection points that are hard to locate in the raw waveform \cite{elgendi2012}.

\begin{table}[t]
\caption{Common PPG Features and Interpretation Risks}
\label{tab:features}
\centering
\footnotesize
\renewcommand{\arraystretch}{1.15}
\setlength{\tabcolsep}{3pt}

\begin{tabular}{|p{0.24\columnwidth}|p{0.31\columnwidth}|p{0.36\columnwidth}|}
\hline
\textbf{Feature} & \textbf{Intended meaning} & \textbf{Major risk} \\
\hline
Systolic amplitude & Pulse volume & Contact/optical variation \\
\hline
Upstroke slope & Ejection and stiffness & Motion and filtering effects \\
\hline
Rise time & Transit-related behavior & Landmark instability \\
\hline
Pulse width & Waveform duration & Heart-rate dependence \\
\hline
Dicrotic notch & Reflection/valve timing & Often absent in wrist PPG \\
\hline
Derivative peaks & Morphological detail & Noise amplification \\
\hline
\end{tabular}
\end{table}

Table~\ref{tab:features} pairs common features with their intended meaning and their main risk. Systolic amplitude reflects pulse volume but responds just as readily to contact pressure and optical coupling. Upstroke slope and rise time depend on precise landmark detection, which motion and filtering disturb. Pulse width varies with heart rate, the dicrotic notch is often absent at the wrist, and derivative peaks amplify noise. Wrist-worn sensors suffer most, because contact and motion change continuously. Filtering, smoothing, baseline correction, and beat rejection make the signal look cleaner, but they can also shift timing landmarks or remove awkward segments, so the features a model sees may describe a better signal than the device actually records.

\section{Taxonomy of Cuffless BP Estimation Methods}
\label{sec:taxonomy}
\begin{figure}[t]
\centering

\resizebox{0.98\columnwidth}{!}{%
\begin{tikzpicture}[
    every node/.style={font=\footnotesize},
    main/.style={
        draw,
        rounded corners=2pt,
        align=center,
        minimum height=7mm,
        text width=2.15cm,
        inner sep=2.5pt
    },
    root/.style={
        draw,
        rounded corners=2pt,
        align=center,
        minimum height=8mm,
        text width=2.4cm,
        inner sep=3pt,
        font=\footnotesize\bfseries
    },
    cross/.style={
        draw,
        rounded corners=2pt,
        align=center,
        minimum height=7mm,
        text width=7.0cm,
        inner sep=2.5pt
    },
    arr/.style={-{Latex[length=1.7mm]}, thick}
]

\node[root] (root) at (0,0)
{Cuffless BP\\estimation};

\node[main] (morph) at (-3.0,-1.5)
{PPG morphology};

\node[main] (timing) at (0,-1.5)
{PAT/PTT timing};

\node[main] (deep) at (3.0,-1.5)
{Deep learning};

\node[main] (multi) at (-1.55,-2.8)
{Multimodal fusion};

\node[main] (personal) at (1.55,-2.8)
{Personalized /\\adaptive models};

\coordinate (bus) at (0,-0.75);

\draw (root.south) -- (bus);

\draw (bus) -- (-3.0,-0.75);
\draw (bus) -- ( 3.0,-0.75);

\draw[arr] (-3.0,-0.75) -- (morph.north);
\draw[arr] ( 0.0,-0.75) -- (timing.north);
\draw[arr] ( 3.0,-0.75) -- (deep.north);

\draw[arr]
    (-1.55,-0.75) -- (-1.55,-2.25) -- (multi.north);

\draw[arr]
    (1.55,-0.75) -- (1.55,-2.25) -- (personal.north);

\node[cross] (validation) at (0,-4.05)
{Validation-aware design, calibration, and evaluation};

\draw[dashed, thick]
    (-3.8,-3.45) -- (3.8,-3.45);

\draw[dashed] (morph.south) -- ++(0,-0.35);
\draw[dashed] (timing.south) -- ++(0,-1.65);
\draw[dashed] (deep.south) -- ++(0,-0.35);
\draw[dashed] (multi.south) -- ++(0,-0.30);
\draw[dashed] (personal.south) -- ++(0,-0.30);

\draw[arr] (0,-3.45) -- (validation.north);

\end{tikzpicture}%
}

\caption{High-level taxonomy of cuffless BP estimation approaches. Validation-aware design applies across all categories.}

\label{fig:taxonomy}
\end{figure}
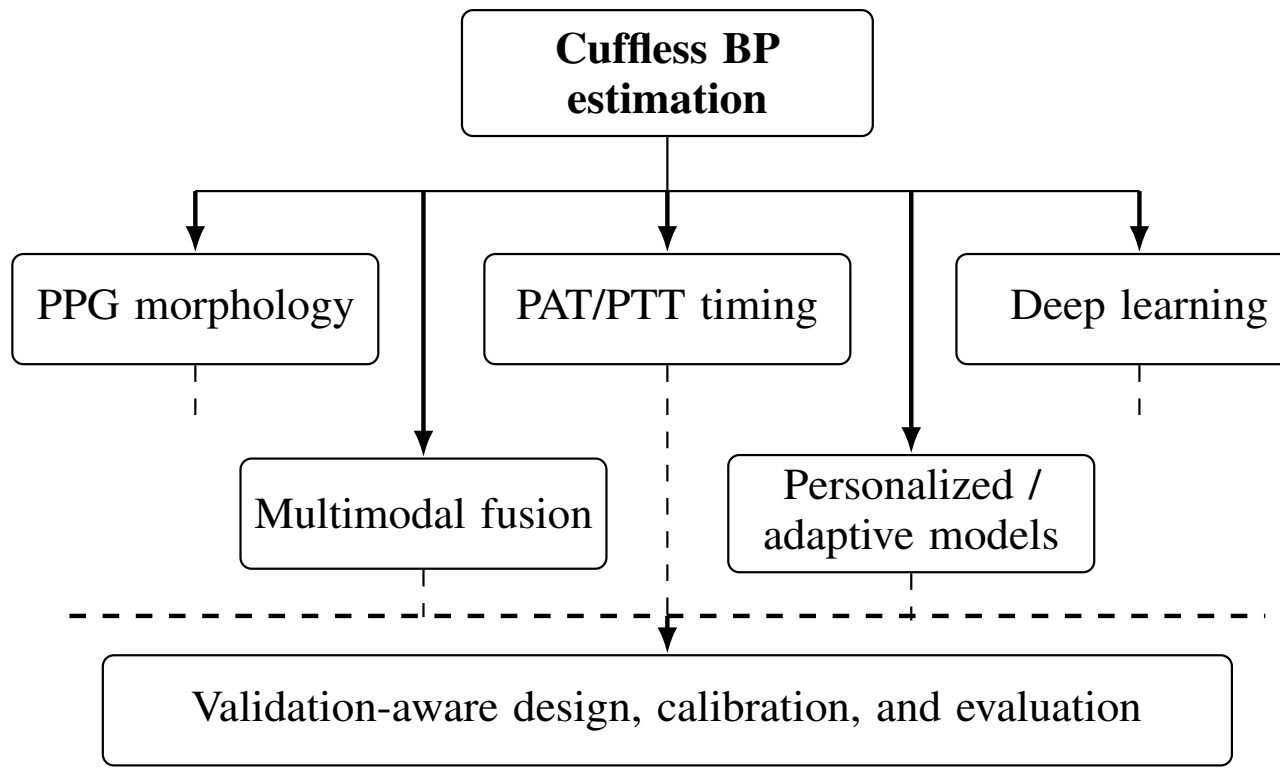

Fig.~\ref{fig:taxonomy} groups the estimation methods. The three upper branches differ in how the input is represented: handcrafted morphology, pulse timing, or representations learned by a deep network. Multimodal fusion and personalized or adaptive models build on these branches, and the dashed layer at the bottom marks validation, calibration, and evaluation design as concerns shared by every category.

\subsection{Feature-Based Regression}
Classical methods apply linear regression, support vector regression, random forests, gradient boosting, or shallow neural networks to handcrafted PPG or ECG--PPG features \cite{kachuee2015,kachuee2017,shin2017,radha2019,shao2020}. They are interpretable and computationally light, but their accuracy is capped by feature quality, and they are sensitive to sensor artifacts, subject variability, and calibration choices. Ensembles such as random forests and gradient boosting cope with noisy or redundant features better than linear regression, at some cost in interpretability. Support vector regression needs careful scaling but performs well with small training sets, a useful property when most datasets contain few subjects. Under subject-independent evaluation, González et al.\ found that feature-based models did best on the two smaller benchmark datasets, whereas deep networks led only on the largest \cite{gonzalez2023}.

\subsection{Deep Learning}
Deep-learning methods learn representations directly from raw or transformed waveforms with convolutional, recurrent, LSTM, transformer, U-Net, or spectro-temporal architectures \cite{slapnicar2019,elhajj2020,elhajj2021,panwar2020,li2020,athaya2021,ibtehaz2022,pan2024,mohammadi2025}. No single architecture has consistently demonstrated superior performance across datasets and evaluation settings. Convolutional networks are efficient and capture local shape but need extra machinery for the long-range dependencies that recurrent and transformer models handle directly, usually at higher computational and data cost. U-Net-style methods such as PPG2ABP reconstruct the full arterial waveform instead of a single SBP/DBP pair. This tends to produce physiologically plausible outputs but makes errors harder to express in the mmHg terms clinicians use \cite{ibtehaz2022}.

Models also differ in how the waveform is presented to them. Some take the raw PPG, some add its derivatives, and some convert it into a time--frequency image, such as a spectro-temporal representation or a continuous wavelet transform, before a convolutional network \cite{slapnicar2019,zhang2026sensors}. Zhang et al.\ compared seven such representations across several architectures \cite{zhang2026sensors}, which treats the representation as a design choice in its own right; it deserves to be reported as carefully as the architecture. Greater capacity also means a greater risk of fitting dataset-specific artifacts, subject identity, or preprocessing choices, and the small number of subjects in most datasets makes that risk worse.

\subsection{Multimodal, Personalized, and Physics-Informed Models}
Multimodal approaches combine PPG with ECG, accelerometry, phonocardiography, demographic variables, or prior cuff readings. The extra inputs can improve robustness, but they add hardware and burden for the user. Personalized models reduce inter-subject variability by adapting to each person; Carlucci et al.\ reported that patient-specific fine-tuning reduced error by up to 64\% \cite{carlucci2026}. Because such adaptation usually requires reference cuff readings, a personalized model's accuracy means little without its calibration requirements.

Physics-informed models take a middle path between hand-built features and unconstrained networks. Sel et al.\ added a Taylor approximation of known, slowly varying cardiovascular relationships to the training of a neural network and reduced the amount of labeled data needed by about fifteen times on average, using bioimpedance signals \cite{sel2023}. The approach has not yet been tested on PPG at scale, but it shows how the physiology described earlier can constrain a model, so that the network does not have to rediscover it from data.

\subsection{On-Device Deployment}
Continuous monitoring is only practical if inference runs on the wearable itself, within the memory and energy budget of a microcontroller. Potail et al.\ demonstrated a TinyML pipeline for continuous BP estimation from PPG on resource-constrained hardware \cite{potail2026}. Burrello et al.\ combined hardware-aware neural architecture search with quantization to fit PPG-to-BP models onto the GAP8 ultra-low-power processor; every optimized model fit in memory, whereas the most accurate reference network on the largest dataset did not \cite{burrello2024}. An extended version of that work reported up to 7.5 times fewer parameters at comparable or lower error and estimated that processing energy alone would allow weeks of continuous monitoring on a small battery \cite{carlucci2026}. Khurana et al.\ examined practical factors such as sampling frequency for deep-learning-based estimation \cite{khurana2026}. Compression changes a model, however, and an on-device model must be validated as deployed, at its actual sampling rate and numerical precision, not only in its full-precision form.

\subsection{Emerging Directions: Remote PPG and Federated Learning}
Remote photoplethysmography (rPPG) extracts a PPG-like signal from video of the face or skin and removes the contact sensor altogether. Schrumpf et al.\ applied BP models trained on contact PPG to rPPG from facial video and reported reduced but non-trivial accuracy, stating the performance drop explicitly \cite{schrumpf2021}. rPPG inherits every calibration and dataset-shift problem of contact PPG and adds lighting, skin-tone-dependent signal quality, compression artifacts, and head motion.

Federated learning addresses a different obstacle: privacy rules make waveform data hard to pool across institutions. Brophy et al.\ trained a generative model for continuous arterial waveform estimation from PPG in a federated setting, in which simulated client devices share model updates instead of raw data \cite{brophy2021}. Such training could reach larger and more diverse populations without centralizing sensitive recordings. Both approaches remain under evaluation, and current evidence is insufficient to establish improved BP accuracy.

\subsection{Comparative Summary}
Table~\ref{tab:approaches} compares these approaches by input, main strength, main limitation, and calibration dependence. The approaches differ substantially in their calibration requirements, input signals, and evaluation conditions, which makes direct comparison of reported accuracy difficult. Pulse-timing and personalized methods rely on reference readings explicitly. Deep-learning methods are often presented as calibration-free, a claim that holds only under the evaluation conditions examined in the next section. The newer directions change how data are acquired, processed, or shared, but they leave the measurement problem itself in place: a remote, federated, or on-device model must meet the same validation requirements as any other.

\begin{table*}[t]
\caption{Comparison of Cuffless BP Estimation Approaches}
\label{tab:approaches}
\centering
\footnotesize
\renewcommand{\arraystretch}{1.15}
\setlength{\tabcolsep}{4pt}

\begin{tabularx}{\textwidth}{|
L{0.15\textwidth}|
L{0.17\textwidth}|
L{0.22\textwidth}|
L{0.22\textwidth}|
>{\RaggedRight\arraybackslash}X|}
\hline
\textbf{Approach} & \textbf{Typical input} & \textbf{Main strength} & \textbf{Main limitation} & \textbf{Calibration dependence} \\
\hline
Feature-based regression & Handcrafted PPG or ECG--PPG features & Interpretable, low computational cost & Bounded by feature quality; sensitive to artifacts & Sensitive to calibration choices \\
\hline
Pulse timing (PAT/PTT) & ECG and one or two PPG sites & Physiological link to arterial stiffness & PEP, vascular tone, and aging confound timing & Generally requires calibration \\
\hline
Deep learning & Raw or transformed waveforms & Captures nonlinear waveform patterns & Overfitting; limited interpretability & Often reported as calibration-free; vulnerable to leakage \\
\hline
Waveform reconstruction & PPG mapped to the arterial waveform & Physiologically plausible output shape & Errors harder to express in mmHg & Depends on evaluation design \\
\hline
Multimodal fusion & PPG with ECG, accelerometry, PCG, or demographics & Potentially more robust & Device complexity and user burden & Varies by design \\
\hline
Personalized or adaptive & Population model plus user data & Reduces inter-subject variability & Requires reference readings & Explicitly calibration-dependent \\
\hline
Remote PPG & Facial or skin video & No contact sensor & Lighting, skin tone, compression, and head motion & Inherits contact-PPG requirements \\
\hline
Federated training & Distributed waveform data & Larger populations without pooling raw data & Does not by itself adapt to new users & Unchanged by the training scheme \\
\hline
Physics-informed & Sensor data plus physiological model & Less labeled data; interpretable structure & Model assumptions must hold for the user & Reduced, not removed \\
\hline
On-device (TinyML) & Compressed versions of the above & Low power; local processing & Accuracy loss from compression; small memory & Same as the source model \\
\hline
\end{tabularx}

\end{table*}

\section{Machine Learning Pitfalls}
\label{sec:mlpitfalls}
Cuffless BP models are usually scored by MAE, RMSE, ME, and SD. These metrics are meaningful only when supported by an appropriate experimental design. Several recurring issues can substantially affect the reported results.

A major concern is \emph{subject leakage}. Long recordings are cut into many short, overlapping or adjacent windows \cite{wang2023pulsedb,sanches2024}. When windows from the same person end up in both the training and the test set, the model can recognize persistent subject-specific properties (vascular anatomy, sensor characteristics, individual physiology) instead of learning a relationship that holds for new people. The impact of subject leakage can be substantial. In a benchmark of public datasets, González et al.\ found that leaky random splits made models appear far better than subject-independent splits, while under subject-independent evaluation most models improved only modestly on a naive predictor that always outputs the training-set mean, and some did not beat it at all \cite{gonzalez2023}. Windowing also inflates apparent dataset size, since many windows may come from few people, so studies should report the number of subjects, not only the number of windows, in each partition.

\begin{figure}[t]
\centering
\begin{tikzpicture}[
  every node/.style={font=\footnotesize},
  tr/.style={draw, fill=blue!25, minimum width=4mm, minimum height=3.2mm, inner sep=0pt},
  te/.style={draw, fill=orange!60, minimum width=4mm, minimum height=3.2mm, inner sep=0pt}
]
\node[anchor=west] at (-0.2,2.55) {\textbf{(a) Record-wise split}};
\foreach \s/\y in {1/2.0,2/1.5,3/1.0,4/0.5}{
  \node[anchor=east] at (0.55,\y) {S\s};
  \foreach \w in {0,...,5}{
    \pgfmathtruncatemacro{\t}{mod(\w+\s,3)==0 ? 1 : 0}
    \ifnum\t=1
      \node[te] at (0.85+\w*0.45,\y) {};
    \else
      \node[tr] at (0.85+\w*0.45,\y) {};
    \fi
  }
}
\node[anchor=west] at (4.1,2.55) {\textbf{(b) Subject-wise split}};
\foreach \s/\y in {1/2.0,2/1.5,3/1.0}{
  \node[anchor=east] at (4.85,\y) {S\s};
  \foreach \w in {0,...,5}{ \node[tr] at (5.15+\w*0.45,\y) {}; }
}
\node[anchor=east] at (4.85,0.5) {S4};
\foreach \w in {0,...,5}{ \node[te] at (5.15+\w*0.45,0.5) {}; }
\node[tr] at (1.6,-0.15) {}; \node[anchor=west] at (1.75,-0.15) {training window};
\node[te] at (4.6,-0.15) {}; \node[anchor=west] at (4.75,-0.15) {test window};
\end{tikzpicture}
\caption{Record-wise and subject-wise partitioning of windowed recordings from four subjects (S1--S4). In (a), windows from every subject appear in both sets, allowing subject leakage; in (b), the test subject is never seen during training.}
\label{fig:splits}
\end{figure}
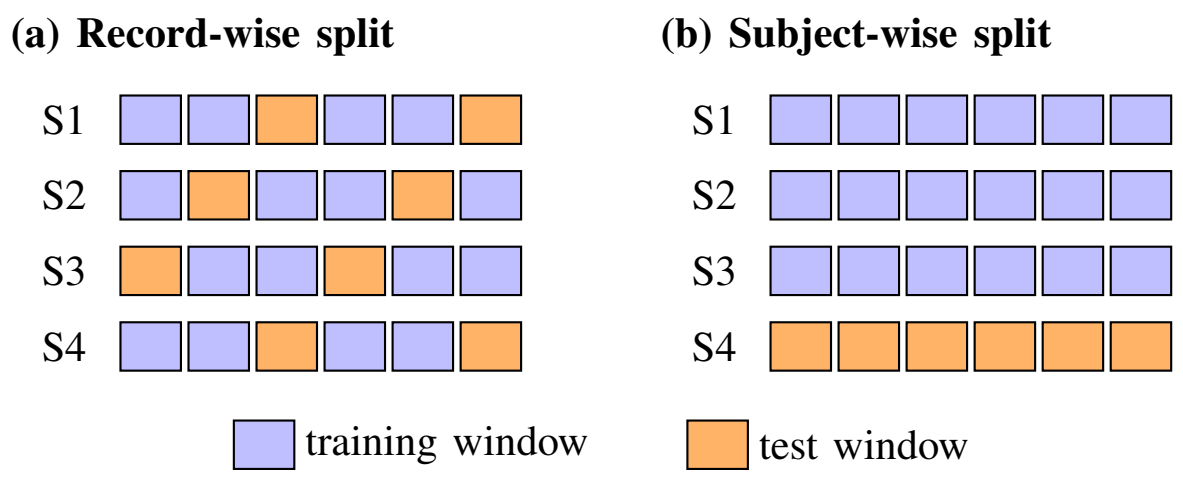

Fig.~\ref{fig:splits} contrasts the two ways of partitioning such data. In a record-wise split, windows from every subject are spread across both sets, and the test set contains no new individual. In a subject-wise split, such as leave-one-subject-out or subject-grouped cross-validation, every window of a test subject is withheld from training. Subject-wise partitioning is the minimum requirement for any claim that a model works on new users.

\emph{Temporal leakage} is subtler. Neighboring windows from one recording share morphology, heart rate, and vascular state, so splitting them at random produces a test set much closer to the training data than any deployment would. Training on earlier portions of a recording and testing on later ones avoids the problem and gives a first look at how performance holds up over time. Leakage can also enter through preprocessing: normalization statistics, feature selection, or signal-quality thresholds computed on the full dataset carry information from the test set into training, so each of these steps should be fitted on the training data alone.

\emph{Imbalance} is the third problem. Public datasets contain far more normotensive or stable samples than hypertensive or hypotensive ones, so a model can post a low average error while failing in exactly the ranges where treatment decisions are made. Evaluating only clean, motionless segments hides the effect of motion and low perfusion in the same way. Finally, a subject-wise split still tests generalization only within one dataset, where sensors, protocols, and populations are shared. External validation on independently collected data is the stronger test, and few studies report it. In all of these cases the reported error describes the test design as much as the model, which is why validation practice is the subject of the next section.

\section{Validation: Evidence Types, Metrics, and Standards}
\label{sec:validation}
Evidence in this field comes in several kinds, and they do not support the same conclusions. Table~\ref{tab:evidence} lists the main types with what each contributes, what it supports, and its main limitation. Physiological and PTT/PAT studies explain why a signal may carry BP-related information but say nothing about device accuracy. Classical and deep-learning studies show algorithmic feasibility, but their results depend on the dataset split, feature quality, and overfitting. Dataset and validation papers make benchmarking possible, while leaving the realism of the data and the validation of cuffless devices unresolved.

\begin{table*}[t]
\caption{Major Evidence Categories in PPG-Based Cuffless BP Research}
\label{tab:evidence}
\centering
\footnotesize
\renewcommand{\arraystretch}{1.15}
\setlength{\tabcolsep}{4pt}

\begin{tabularx}{\textwidth}{|
L{0.17\textwidth}|
L{0.27\textwidth}|
L{0.25\textwidth}|
>{\RaggedRight\arraybackslash}X|}
\hline
\textbf{Evidence type} & \textbf{Typical contribution} & \textbf{What it supports} & \textbf{Main limitation} \\
\hline
PPG physiology & Explains optical waveform morphology and fiducial features & Supports interpretation of PPG-derived features & Does not establish BP accuracy \\
\hline
PTT/PAT theory & Links pulse timing to vascular stiffness and pulse propagation & Supports timing-based BP estimation mechanisms & Sensitive to PEP, tone, stiffness, and calibration \\
\hline
Classical ML & Uses engineered waveform and timing features for regression & Demonstrates algorithmic feasibility & Strongly affected by dataset split and feature quality \\
\hline
Deep learning & Learns waveform representations from raw or transformed signals & Captures nonlinear patterns & Risk of overfitting and limited interpretability \\
\hline
Datasets and validation & Provides waveform resources and evaluation standards & Supports benchmarking and clinical translation discussion & Dataset realism and cuffless validation remain unresolved \\
\hline
\end{tabularx}

\end{table*}

\begin{table*}[!t]

\caption{Representative Studies Positioned on the Evidence Ladder}
\label{tab:methodcomparison}

\centering
\footnotesize
\renewcommand{\arraystretch}{1.12}
\setlength{\tabcolsep}{2pt}

\begin{tabular}{
|>{\raggedright\arraybackslash}p{0.13\textwidth}
|>{\raggedright\arraybackslash}p{0.13\textwidth}
|>{\raggedright\arraybackslash}p{0.16\textwidth}
|>{\raggedright\arraybackslash}p{0.16\textwidth}
|>{\raggedright\arraybackslash}p{0.20\textwidth}
|>{\raggedright\arraybackslash}p{0.16\textwidth}|
}

\hline

\textbf{Study} &
\textbf{Signal / Model} &
\textbf{Dataset (subjects)} &
\textbf{Subject-independent?} &
\textbf{Reported performance} &
\textbf{Evidence-ladder tier} \\

\hline

Kachuee et al., 2017 \cite{kachuee2017} &
PPG + ECG (PAT); AdaBoost &
UCI/MIMIC-II; 942 subjects &
Not reported &
SBP 11.17 mmHg; DBP 5.35 mmHg
(corroborated in \cite{wang2020cvprw,ibtehaz2022}) &
Algorithmic feasibility \\

\hline

Slap\v{n}i\v{c}ar et al., 2019 \cite{slapnicar2019} &
PPG + derivatives; spectro-temporal CNN &
MIMIC-III; 510 subjects &
Yes; leave-one-subject-out &
SBP 9.43 mmHg; DBP 6.88 mmHg &
Validated measurement performance
(single institution) \\

\hline

Ibtehaz et al., 2022 \cite{ibtehaz2022} &
PPG only; U-Net waveform reconstruction (PPG2ABP) &
UCI/MIMIC-II; 942 subjects
(same source pool as Kachuee et al.) &
No; random-window split permits subject overlap
\cite{mehta2024} &
SBP $5.73 \pm 9.16$ mmHg;
DBP $3.45 \pm 6.15$ mmHg &
Algorithmic feasibility
(leakage confirmed) \\

\hline

Pan et al., 2024 \cite{pan2024} &
PPG only; waveform reconstruction &
VitalDB (extended) &
Not reported &
SBP $2.16 \pm 1.53$ mmHg;
DBP $1.12 \pm 0.59$ mmHg &
Algorithmic feasibility \\

\hline

Mohammadi et al., 2025 \cite{mohammadi2025} &
PPG only; hybrid CNN--BiLSTM with attention &
Not fully disclosed &
Not reported;
5-fold cross-validation &
SBP 1.88 mmHg;
DBP 1.34 mmHg &
Algorithmic feasibility \\

\hline

Zhang et al., 2026 \cite{zhang2026sensors} &
PPG only; CWT + CNN &
UCI Open BP Database;
external validation on 26 participants &
Partial; independent participant group used for external validation &
Source dataset:
SBP $3.80 \pm 5.02$ mmHg;
DBP $1.65 \pm 2.70$ mmHg.

External validation:
$R=0.96$ (SBP);
$R=0.74$ (DBP) &
Algorithmic feasibility with limited external validation \\

\hline

\end{tabular}

\end{table*}

We order these kinds of evidence as a ladder with four tiers: \emph{physiological plausibility} (a signal correlates with BP), \emph{algorithmic feasibility} (a model reproduces BP on a specific dataset), \emph{validated measurement performance} (the model generalizes under subject-independent, temporally separated, and calibration-transparent evaluation), and \emph{clinical utility} (the resulting measurement changes a diagnostic or treatment decision safely). A low error on a retrospective database reaches the second tier and no further. It shows that an algorithm reproduces the reference labels of that database, not that the relationship holds for unseen subjects, sensors, or physiological states.

\subsection{Representative Studies Positioned on the Evidence Ladder}
Table~\ref{tab:methodcomparison} places six representative studies on the ladder, covering feature-based regression, convolutional deep learning, waveform reconstruction, and recent 2026 work. They were chosen because each reports enough methodological detail to be positioned with confidence and because a single, unambiguous source for its reported error could be identified. All error values come from the cited papers; for Kachuee et al., the same values are also reported by two later independent papers \cite{wang2020cvprw,ibtehaz2022}.

Kachuee et al.\ and Ibtehaz et al.\ draw on the same subject pool, the UCI Cuff-Less Blood Pressure Estimation Dataset, a 942-patient MIMIC-II subset compiled by Kachuee et al., so their results compare modeling approaches on identical data and say nothing about other populations \cite{kachuee2017,ibtehaz2022}. Kachuee et al.\ report a general, calibration-free model without a subject-independent split, which places the study at algorithmic feasibility \cite{kachuee2017}. Slap\v{n}i\v{c}ar et al.\ use leave-one-subject-out evaluation and meet the subject-independence criterion for validated measurement performance, although their single-institution dataset leaves external validity untested \cite{slapnicar2019}. For Ibtehaz et al., a later inspection of the published source code found that PPG2ABP splits overlapping 10-second windows into training and test sets at random, so windows from the same patient appear in both, a confirmed case of subject leakage \cite{mehta2024}. Pan et al.\ and Mohammadi et al.\ do not report subject-independent splits, so despite their low errors both remain at algorithmic feasibility until independently replicated \cite{pan2024,mohammadi2025}. Zhang et al.\ are the only group to add an external test: after training on a public dataset, they compared their model with an oscillometric cuff in a separate group of 26 participants \cite{zhang2026sensors}. That is stronger evidence than a retrospective split, although the external results are reported only as correlations.

The numbers in Table~\ref{tab:methodcomparison} also show how reporting has drifted. The two earliest studies report SBP errors of roughly 9--11\,mmHg, the most recent about 2\,mmHg. The lowest errors, however, come from studies without a reported subject-independent split, and Mohammadi et al.\ do not fully disclose their dataset. Zhang et al.\ report external correlations of $R=0.96$ for SBP and $R=0.74$ for DBP. A high correlation shows that estimates move with the reference; it does not show that they agree with it in mmHg, which is what Bland--Altman analysis and the validation standards test. The decline in reported MAE from 2017 to 2025 should not be read as steady progress, because the studies differ in dataset, population, preprocessing, and validation protocol, and the study with confirmed leakage sits in the middle of the range. None of the six reaches clinical utility, since none reports an ambulatory or outcome-linked validation.

\subsection{Agreement and BP-Range Stratification}
\begin{figure}[t]
\centering
\begin{tikzpicture}[every node/.style={font=\footnotesize}, x=0.62mm, y=1.5mm]
\draw[-{Latex[length=1.6mm]}] (85,-16) -- (182,-16);
\draw[-{Latex[length=1.6mm]}] (85,-16) -- (85,15);
\node[below=11pt] at (133,-16) {Mean of device and reference BP (mmHg)};
\node[rotate=90] at (69,0) {Device $-$ reference (mmHg)};
\foreach \x in {100,120,140,160} { \draw (\x,-16) -- (\x,-15.2); \node[below] at (\x,-16) {\x}; }
\foreach \y in {-10,0,10} { \draw (85,\y) -- (86.2,\y); \node[left] at (85,\y) {\y}; }
\draw[thick] (85,-1.5) -- (172,-1.5) node[right]{mean};
\draw[dashed] (85,7.5) -- (172,7.5) node[right]{$+1.96\,$SD};
\draw[dashed] (85,-10.5) -- (172,-10.5) node[right]{$-1.96\,$SD};
\foreach \x/\y in {92/2,96/-1,99/3,103/0,106/-2,108/4,111/1,114/-1,117/2,119/-3,122/0,125/-2,127/1,130/-4,133/-2,136/-5,139/-3,142/-6,145/-4,148/-8,151/-6,154/-9,158/-8,162/-11,165/-10,168/-13}
  { \fill (\x,\y) circle (0.9pt); }
\end{tikzpicture}
\caption{Schematic Bland--Altman plot. Errors are centered near zero at normal BP but become increasingly negative at high BP, a proportional underestimation that an average error would conceal.}
\label{fig:blandaltman}
\end{figure}
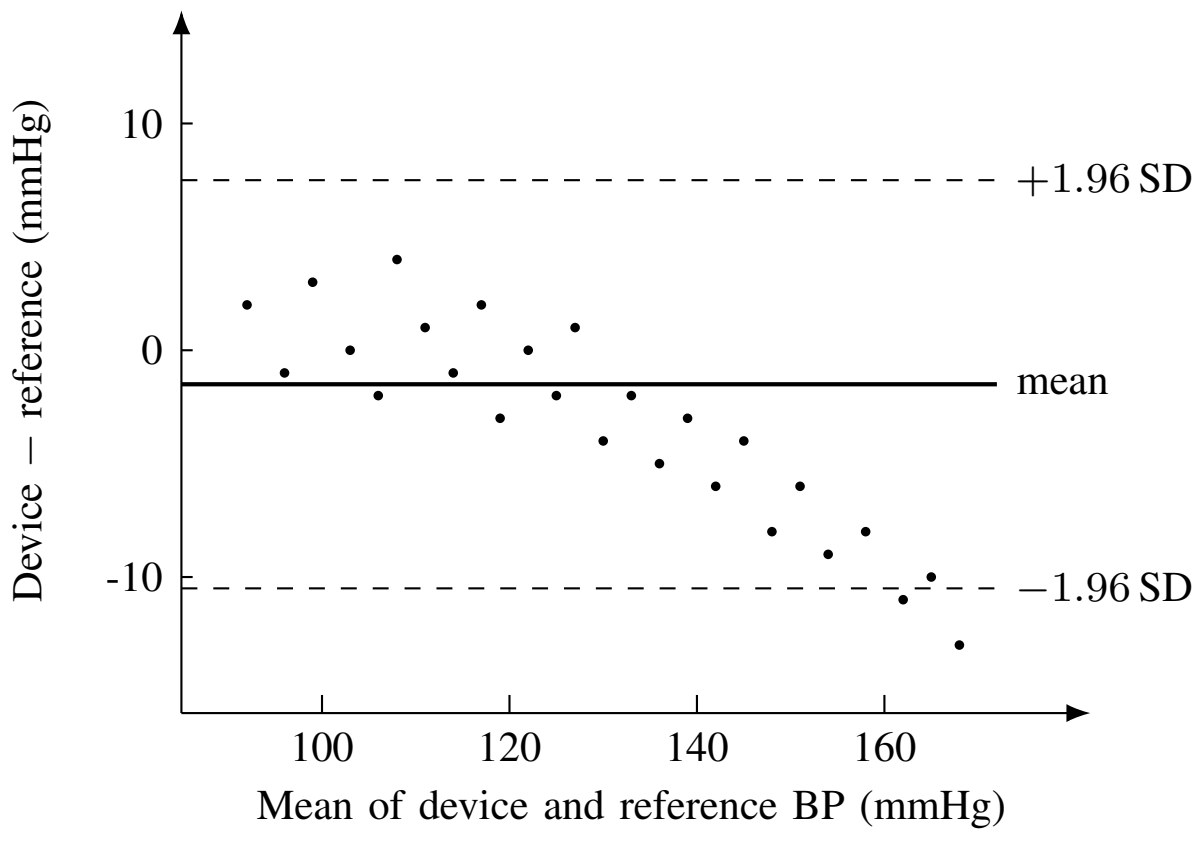

Summary statistics can hide the errors that matter clinically. Mean error can be small when positive and negative errors cancel, and MAE and RMSE average over all samples, so a poorly served subgroup barely moves them. Bland--Altman analysis instead plots the difference between a method and its reference against their mean \cite{bland1986}. Fig.~\ref{fig:blandaltman} shows a schematic example. The mean difference indicates systematic bias, and the limits of agreement, the mean difference $\pm 1.96$ standard deviations, show the range within which most errors fall. In the example, errors center near zero at normal BP but become increasingly negative at high BP. A device with this pattern underestimates the very readings that drive treatment and may be unsafe despite a low average error. Because public datasets contain more normal-range than hypertensive or hypotensive samples, careful studies report performance separately for low, normal, elevated, and high BP whenever the data allow \cite{stergiou2018_universal,hu2023}.

\subsection{Cuff-Based Validation Standards}
The AAMI/ESH/ISO universal standard and ISO 81060-2 define how intermittent BP devices should be validated \cite{stergiou2018_universal,iso81060}, and the oscillometric devices that serve as reference for most cuffless studies are themselves well characterized \cite{sharman2023}. These standards judge agreement in mmHg, commonly requiring a mean difference within approximately 5\,mmHg and a standard deviation not exceeding approximately 8\,mmHg for SBP and DBP, a more clinically meaningful criterion than percentage accuracy. Many algorithmic studies report only MAE, which cannot be checked against these criteria; reporting the mean error and its standard deviation as well costs nothing. ISO 81060-3 extends clinical investigation to continuous non-invasive devices, covering both trending and absolute-accuracy claims, but it explicitly does not assess motion artifacts \cite{iso81060_3}. None of these documents addresses everything a wearable cuffless device faces: continuous wear, calibration dependence, and sensitivity to posture and motion.

\begin{table*}[!t]
\caption{Accuracy and Validation Requirements for Cuffless BP Studies}
\label{tab:requirements}
\centering
\footnotesize
\renewcommand{\arraystretch}{1.15}
\setlength{\tabcolsep}{4pt}

\begin{tabularx}{\textwidth}{|
L{0.18\textwidth}|
L{0.27\textwidth}|
L{0.24\textwidth}|
>{\RaggedRight\arraybackslash}X|}
\hline
\textbf{Requirement} & \textbf{What should be reported} & \textbf{Why it matters} & \textbf{Risk if missing} \\
\hline
Mean error and SD & Bias and spread of error in mmHg & Aligns with BP validation conventions & Average performance may be misleading \\
\hline
MAE and RMSE & Absolute and squared-error summaries & Quantifies prediction error magnitude & Large individual errors may be hidden \\
\hline
Bland--Altman analysis & Agreement and limits of agreement across BP range & Reveals systematic bias and outliers & Unsafe errors may be missed \\
\hline
BP-range stratification & Error in low, normal, elevated, and high BP ranges & Tests clinically important ranges & Hypertension performance may be weak \\
\hline
Subject-independent split & No subject overlap between training and testing & Prevents identity leakage & Inflated benchmark accuracy \\
\hline
Temporal separation & Prevents adjacent windows from crossing train/test sets & Avoids correlation leakage & Unrealistically low error \\
\hline
Calibration protocol & Initial, repeated, or calibration-free design & Defines true device dependency & Cuff dependence may be hidden \\
\hline
Time since calibration & Accuracy after hours, days, or weeks & Tests drift & Short-term accuracy may not persist \\
\hline
Motion and posture & Static, walking, exercise, sleep, and body position & Tests wearable realism & Device may work only in ideal conditions \\
\hline
Failure rate & Rejected or unavailable readings & Prevents silent exclusion bias & Accuracy may reflect only clean segments \\
\hline
\end{tabularx}

\end{table*}

\subsection{Cuffless-Specific Validation}
Mukkamala et al.\ argued that the universal standard is inappropriate for cuffless devices and showed, with examples from the literature and simulations, that many accuracy studies use methods that can produce misleading conclusions \cite{mukkamala2021}. IEEE 1708 covers wearable cuffless devices across measurement modes: short-term, long-term, snapshot, continuous, beat-to-beat, and BP variability \cite{ieee1708,ieee1708a}. The European Society of Hypertension recommends validation tailored to device function and calibration \cite{stergiou2023}, and Hu et al.\ call for frameworks that account for calibration-dependent operation and real-world use \cite{hu2023}.

Table~\ref{tab:requirements} turns these recommendations into a reporting list, giving for each item what should be reported, why it matters, and the risk when it is missing. The items fall into four groups: error statistics (ME and SD, MAE and RMSE, Bland--Altman analysis, BP-range stratification), data separation (subject-independent and temporally separated splits), calibration (protocol and time since calibration), and conditions of use (motion, posture, and failure rate). Leaving out any group hides a specific weakness, from accuracy inflated by leakage to accuracy that reflects only clean segments.

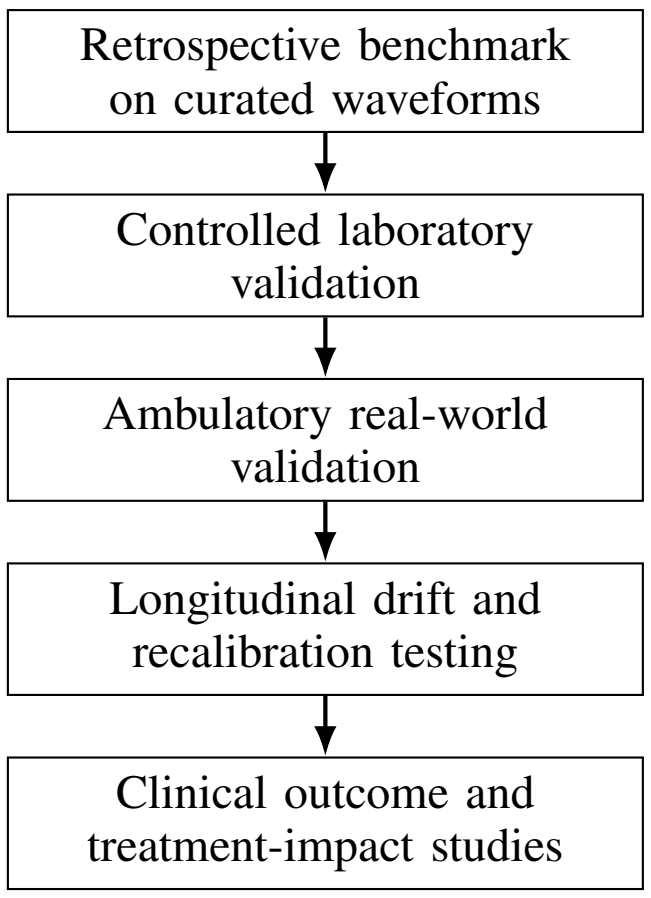
\begin{figure}[t]
\centering
\begin{tikzpicture}[
  node distance=4mm,
  every node/.style={font=\small},
  box/.style={draw, align=center, minimum width=4.2cm, minimum height=8mm},
  arr/.style={-{Latex[length=2mm]}, thick}
]
\node[box] (b1) {Retrospective benchmark\\on curated waveforms};
\node[box, below=of b1] (b2) {Controlled laboratory\\validation};
\node[box, below=of b2] (b3) {Ambulatory real-world\\validation};
\node[box, below=of b3] (b4) {Longitudinal drift and\\recalibration testing};
\node[box, below=of b4] (b5) {Clinical outcome and\\treatment-impact studies};
\draw[arr] (b1) -- (b2);
\draw[arr] (b2) -- (b3);
\draw[arr] (b3) -- (b4);
\draw[arr] (b4) -- (b5);
\end{tikzpicture}
\caption{Validation ladder for cuffless BP technologies. Many studies remain at the retrospective benchmark stage, whereas clinical adoption requires real-world, longitudinal, and outcome-linked evidence.}
\label{fig:ladder}
\end{figure}

Fig.~\ref{fig:ladder} recasts the evidence ladder as five practical validation stages: retrospective benchmarks on curated waveforms, controlled laboratory validation, ambulatory validation in daily life, longitudinal testing of drift and recalibration, and clinical outcome and treatment-impact studies. Retrospective benchmarks are useful for development, but clinical adoption needs evidence from the later stages, and many published studies have not moved past the first.

\begin{table*}[!t]
\caption{Dataset Considerations for Cuffless BP Estimation}
\label{tab:datasets}
\centering
\footnotesize
\renewcommand{\arraystretch}{1.10}
\setlength{\tabcolsep}{3.5pt}

\begin{tabularx}{\textwidth}{|
>{\raggedright\arraybackslash}p{0.15\textwidth}|
>{\raggedright\arraybackslash}p{0.12\textwidth}|
>{\raggedright\arraybackslash}p{0.20\textwidth}|
>{\raggedright\arraybackslash}p{0.11\textwidth}|
>{\raggedright\arraybackslash}p{0.18\textwidth}|
>{\raggedright\arraybackslash}X|}
\hline
\textbf{Dataset} & \textbf{Typical setting} & \textbf{Subjects} & \textbf{Sampling rate} & \textbf{Useful signals} & \textbf{Main limitation} \\
\hline
MIMIC-III Waveform Database & ICU & $\sim$30{,}000 patients (10{,}282 in the clinically matched subset) & 125\,Hz & ECG, PPG, arterial BP & ICU population and care-context bias \cite{johnson2016} \\
\hline
VitalDB & Operating room & 6{,}388 surgical patients & Up to 500\,Hz & Multiple vital signs and waveforms & Anesthesia and OR-specific physiology \cite{lee2022} \\
\hline
PulseDB & Curated benchmark from MIMIC-III/VitalDB & 5{,}361 subjects (5{,}245{,}454 10-s segments) & 125\,Hz & PPG, ECG, arterial BP labels & Inherits ICU/OR setting limitations \cite{wang2023pulsedb} \\
\hline
PPG-BP & Seated, clinical room & 219 subjects aged 20--89 (657 records) & 1\,kHz & Fingertip PPG (2.1-s segments), cuff BP & Very short single-session recordings \cite{liang2018} \\
\hline
Aurora-BP & Laboratory and 24-h ambulatory & More than 1{,}000 participants & Device-dependent & Wrist PPG, ECG, tonometry, accelerometry, ambulatory cuff & Intermittent cuff reference; single device family \cite{mieloszyk2022} \\
\hline
Other ambulatory wearable datasets & Daily life & Varies by study; often small & Device-dependent, commonly 25--100\,Hz & PPG, accelerometry, cuff references & Realistic deployment conditions but harder to collect and standardize \\
\hline
\end{tabularx}
\end{table*}

\section{Datasets and Generalization Limitations}
\label{sec:datasets}
Public waveform datasets have driven much of the progress in cuffless BP research, because they provide synchronized signals and reference labels at a scale few laboratories could collect. MIMIC-III, VitalDB, PulseDB, MIMIC-BP, and PhysioNet resources are the most widely used \cite{goldberger2000,johnson2016,lee2022,wang2023pulsedb,sanches2024}. Table~\ref{tab:datasets} compares them with two datasets recorded outside intensive care. MIMIC-III contains about 30{,}000 ICU patients recorded at 125\,Hz, VitalDB 6{,}388 surgical patients recorded at up to 500\,Hz, and PulseDB 5{,}361 subjects and more than five million 10-s segments curated from those two sources. PulseDB and MIMIC-BP were built to standardize segment selection and labeling, which makes results comparable when studies share subjects and partitions, but the recordings still come from hospital patients. PPG-BP, by contrast, contains 219 seated participants aged 20 to 89 with cuff references, but only three 2.1-second fingertip segments per person \cite{liang2018}. Aurora-BP is the largest ambulatory resource we found: more than 1{,}000 participants wore wrist devices recording PPG, ECG, tonometry, and accelerometry, with reference readings taken in the laboratory and by an ambulatory cuff over 24 hours \cite{mieloszyk2022}.

\begin{figure}[t]
\centering
\includegraphics[width=\linewidth]{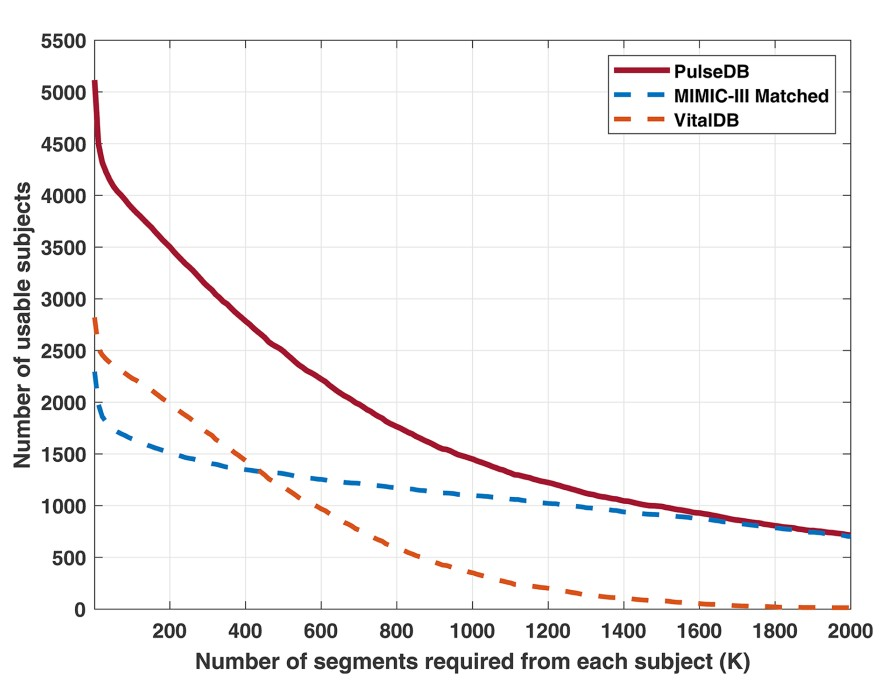}
\caption{Subject coverage of PulseDB, MIMIC-III Matched, and VitalDB as the minimum number of segments per subject increases. Reproduced from \cite{wang2023pulsedb}.}
\label{fig:subjectcoverage}
\end{figure}

How much each subject contributes matters as much as how many subjects there are. Fig.~\ref{fig:subjectcoverage} shows how many subjects remain in PulseDB, MIMIC-III Matched, and VitalDB as the minimum number of segments required per subject increases. At the lowest threshold PulseDB holds about 5{,}000 usable subjects, roughly the sum of its two sources. VitalDB subjects then drop out quickly: the VitalDB curve falls below MIMIC-III Matched at about 400 segments per subject and approaches zero by 2{,}000, where PulseDB converges to MIMIC-III Matched at about 700 subjects. The subjects with long recordings, who tend to dominate windowed training sets, come mainly from the ICU. ICU and operating-room patients are monitored, often sedated or hemodynamically unstable, and recorded with clinical-grade sensors; a wearable must work during walking, sleep, exercise, temperature change, and variable contact. The reference labels raise the same issue, since invasive arterial lines are used only in critically ill, anesthetized, or surgical patients.

\subsection{Evidence on Generalization}
Several recent studies have measured what happens when models leave the data they were trained on. Weber-Boisvert et al.\ trained models on MIMIC and tested them on PPG-BP, and vice versa; in both directions the models lost nearly all predictive power. Features that correlated strongly with SBP in PPG-BP correlated weakly in MIMIC, and the two datasets differed significantly in 20 of 31 pulse features after adjustment for heart rate \cite{weberboisvert2023}. Moulaeifard et al.\ benchmarked deep models across datasets and found in-distribution performance to be an overly optimistic guide: without calibration, their best model reached 13.9/8.5\,mmHg (SBP/DBP) in-distribution and 10.0--18.6/5.9--10.3\,mmHg on external datasets. Errors tracked the difference in BP distributions between training and test data, and models trained on MIMIC generalized poorly \cite{moulaeifard2025}. In the ambulatory Aurora-BP data, Shen et al.\ reported mean absolute differences of about 14\,mmHg for SBP and 10\,mmHg for DBP, by day and by night \cite{shen2025}. These figures are several times the errors that dominate the benchmark literature. Strong performance on hospital data demonstrates algorithmic feasibility; it does not demonstrate readiness for ambulatory use.

\section{Calibration, Personalization, and Drift}
\label{sec:calibration}
Most cuffless systems do not infer absolute BP from PPG or timing signals alone \cite{sharman2020,stergiou2022,mukkamala2015}. They use one or more cuff readings to map the measured signal onto the user's BP. Calibration improves short-term accuracy but leaves the device partly dependent on a cuff, and the degree of that dependence is central to what the device can claim.

\subsection{Why Personalization Is Often Needed}
The relationship between PPG features, PAT or PTT, and BP varies with age, arterial stiffness, vascular geometry, medication, autonomic tone, sensor site, and local perfusion \cite{sharman2020,mukkamala2015}. Two people can share similar PPG features but have different BP, and one person's relationship can shift between rest, stress, exercise, and temperature change. Whether a device uses one reference reading, several, or repeated recalibration defines what it actually measures, and the protocol belongs alongside the accuracy figures. The size of the effect is visible in benchmark data: Moulaeifard et al.\ report in-distribution errors of 9.0/5.8\,mmHg with subject-specific calibration against 13.9/8.5\,mmHg without it \cite{moulaeifard2025}.

\subsection{Calibration-Free Claims}
Calibration-free estimation removes the cuff altogether, but it is the hardest claim to validate, because the model must generalize across subjects, sensors, and physiological states without any personal reference. It is also the claim most exposed to leakage. If windows from one person appear in both training and testing, a model can look calibration-free while relying on characteristics of that person it has already seen; González et al.\ note that some earlier work effectively treated such leakage as a form of calibration \cite{gonzalez2023}. Calibration-free results are credible only under strictly subject-independent, temporally separated evaluation.

\subsection{Calibration Drift}
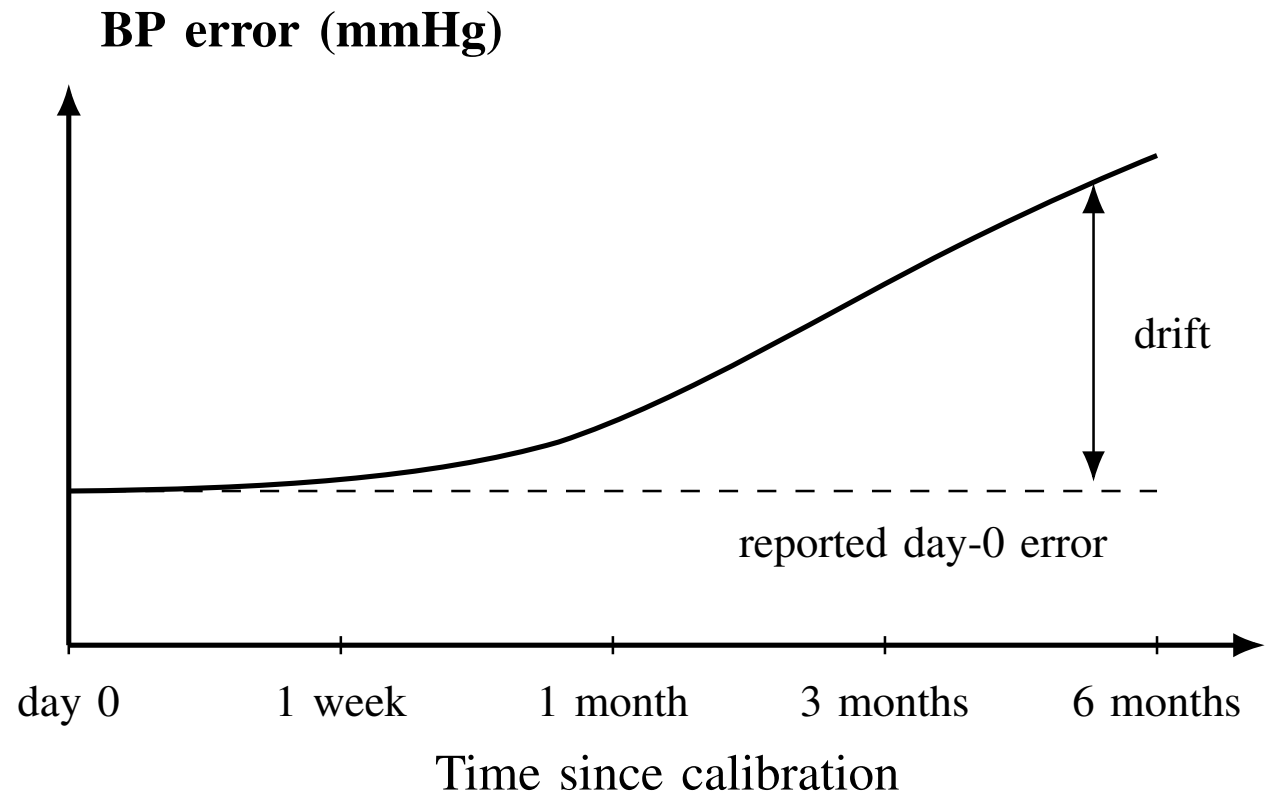
\begin{figure}[t]
\centering

\resizebox{0.95\columnwidth}{!}{%
\begin{tikzpicture}[
    every node/.style={font=\footnotesize}
]

\draw[-{Latex[length=1.8mm]}, thick]
    (0,0) -- (6.6,0);

\draw[-{Latex[length=1.8mm]}, thick]
    (0,0) -- (0,3.1);

\node[
    anchor=south west,
    font=\footnotesize\bfseries
] at (0.05,3.12)
{BP error (mmHg)};

\node[
    anchor=north,
    font=\footnotesize
] at (3.3,-0.48)
{Time since calibration};

\foreach \x/\lbl in {
    0/day 0,
    1.5/1 week,
    3.0/1 month,
    4.5/3 months,
    6.0/6 months
}{
    \draw (\x,-0.05) -- (\x,0.05);

    \node[
        anchor=north,
        font=\scriptsize
    ] at (\x,-0.10) {\lbl};
}

\draw[dashed]
    (0,0.85) -- (6.0,0.85);

\node[
    anchor=north east,
    font=\scriptsize
] at (5.85,0.78)
{reported day-0 error};

\draw[thick]
    (0,0.85)
    .. controls (1.1,0.86) and (2.0,0.92) .. (2.7,1.12)
    .. controls (3.4,1.35) and (4.2,1.85) .. (4.9,2.20)
    .. controls (5.4,2.45) and (5.8,2.62) .. (6.0,2.70);

\draw[<->, >=Latex, thin]
    (5.65,0.90) -- (5.65,2.55);

\node[
    anchor=west,
    font=\scriptsize
] at (5.75,1.72)
{drift};

\end{tikzpicture}%
}

\caption{Schematic illustration of calibration drift. Error measured shortly after calibration may not represent long-term performance.}

\label{fig:drift}
\end{figure}

Even a good calibration does not last indefinitely. Medication changes, disease progression, vascular remodeling, aging, posture, stress, temperature, sensor repositioning, and strap tightness can all alter PPG morphology or pulse timing without any change in central pressure. Fig.~\ref{fig:drift} sketches the consequence. Error stays close to its day-0 value for the first days, then rises over weeks and months, and a study that measures accuracy only during the calibration session sees just the flat start of the curve. A drifting device keeps producing plausible numbers after its calibration has become invalid, which makes drift a safety problem as well as an accuracy problem. Validation should report error as a function of time since calibration, and the recalibration interval should be treated as a core device characteristic \cite{stergiou2022,stergiou2023}.

\subsection{Comparing Approaches to the Drift and Personalization Problem}
The strategies proposed for personalization and drift are not interchangeable. Periodic recalibration against a cuff is the most direct and is implicit in most commercial designs, but it brings back the cuff, and its value depends on how often users actually recalibrate. Feature-level personalization adjusts a population model with a few subject-specific readings, which lightens the calibration burden but must still be evaluated on subjects not used for personalization. Transfer learning and domain adaptation adapt a population model to a new subject, sensor, or population with far less data than training from scratch. Self-supervised pretraining on large amounts of unlabeled PPG offers another route; Pillai et al.'s PaPaGei foundation model demonstrates this for PPG-based physiological signals in general, although its benefit for BP has not been evaluated directly \cite{pillai2024papagei}. Explicit drift detection, which treats drift as a condition to flag instead of a parameter to re-fit, has received little attention.

\begin{table}[!t]
\caption{Calibration Issues in Cuffless BP Monitoring}
\label{tab:calibration}
\centering
\footnotesize
\renewcommand{\arraystretch}{1.10}
\setlength{\tabcolsep}{3pt}

\begin{tabular}{|
>{\raggedright\arraybackslash}p{0.24\columnwidth}|
>{\raggedright\arraybackslash}p{0.35\columnwidth}|
>{\raggedright\arraybackslash}p{0.33\columnwidth}|}
\hline
\textbf{Issue} & \textbf{Reason} & \textbf{Reporting need} \\
\hline
Initial calibration & Aligns model output to individual BP level & Number and type of reference readings \\
\hline
Repeated calibration & Maintains accuracy as physiology changes & Recalibration interval and trigger \\
\hline
Calibration-free design & Avoids cuff dependency & Subject-independent external validation \\
\hline
Calibration drift & Signal--BP relationship changes over time & Error versus time since calibration \\
\hline
Physiological state & Rest, stress, sleep, exercise, and posture alter signals & Calibration and use conditions \\
\hline
\end{tabular}
\end{table}

None of these strategies removes the need for transparent reporting. They differ in where the burden of adaptation falls, not in whether adaptation is needed. Table~\ref{tab:calibration} lists what should be reported for each calibration issue: the number and type of reference readings for initial calibration, the recalibration interval and trigger for repeated calibration, subject-independent external validation for calibration-free designs, error against time since calibration for drift, and the calibration and use conditions when physiological state changes.

\section{Bias, Equity, and Optical Sensing Limitations}
\label{sec:bias}
\begin{table*}[!t]
\caption{Sources of Bias and Generalization Risk in Cuffless BP Estimation}
\label{tab:bias}
\centering
\footnotesize
\renewcommand{\arraystretch}{1.10}
\setlength{\tabcolsep}{4pt}

\begin{tabularx}{\textwidth}{|
>{\raggedright\arraybackslash}p{0.18\textwidth}|
>{\raggedright\arraybackslash}p{0.28\textwidth}|
>{\raggedright\arraybackslash}p{0.28\textwidth}|
>{\raggedright\arraybackslash}X|}
\hline
\textbf{Bias source} & \textbf{How it affects cuffless BP} & \textbf{What should be reported} & \textbf{Clinical risk} \\
\hline
Skin and tissue optics & Alters PPG signal strength and morphology & Skin tone or pigmentation-relevant descriptors when feasible & Uneven accuracy across users \\
\hline
Age and vascular stiffness & Changes pulse wave velocity and morphology & Age range and vascular disease status & Poor transfer across age groups \\
\hline
Medication and disease & Alters vascular tone, heart rate, and BP dynamics & Medication and comorbidity information & Misleading BP trends \\
\hline
Care setting & ICU/OR data differ from daily-life wearable use & Dataset source and acquisition context & Benchmark accuracy may not translate \\
\hline
Sensor site & Wrist, finger, ear, and phone PPG differ in signal quality & Sensor location and attachment method & Site-specific failure modes \\
\hline
Motion and posture & Corrupts PPG and changes hydrostatic conditions & Activity and posture during measurement & Failure during real-world use \\
\hline
\end{tabularx}
\end{table*}

Because PPG is optical, its quality depends on skin and tissue properties, sensor wavelength, contact pressure, perfusion, and motion. Pulse-oximetry studies have documented errors associated with skin pigmentation \cite{bickler2005,sjoding2020,fawzy2022}. For PPG itself the picture is mixed. Melanin absorbs strongly at the green wavelengths most wrist devices use, and several studies have reported larger heart-rate errors in darker skin \cite{fine2021}, whereas Bent et al.\ found no significant effect of skin tone on heart-rate error across six commercial wearables \cite{bent2020}. Heart rate depends only on pulse timing, while BP estimation also relies on amplitude and shape, which may be more sensitive to optical attenuation. The evidence therefore points to a plausible risk for cuffless BP, not a demonstrated bias, and settling the question will take validation across diverse skin tones.

The make-up of training data creates a second source of bias. The main public datasets come from hospitalized adults, so models trained on them may not transfer to healthy users, ambulatory hypertensive patients, older adults, pregnant individuals, pediatric populations, or people with vascular disease, however large the dataset. Studies should report demographic and physiological descriptors (age, sex, BP range, health status, medication use, comorbidities, sensor site, and calibration protocol) where feasible and ethically appropriate, together with skin-tone or pigmentation-relevant variables where available, and should use them to test robustness, not to support unfounded biological assumptions.

Table~\ref{tab:bias} collects six sources of bias (skin and tissue optics, age and vascular stiffness, medication and disease, care setting, sensor site, and motion and posture), with how each affects cuffless BP, what should be reported, and the clinical risk when it is ignored. Several rows tie directly to mechanisms described earlier. Age and vascular stiffness change pulse wave velocity and waveform shape, so a model trained mostly on one age range may transfer poorly to another. Medication and disease alter vascular tone, heart rate, and BP dynamics, so a device can report misleading trends after a change in treatment. Wrist, finger, ear, and phone-based PPG differ in signal quality, and each fails in its own way.

\section{Wearable Deployment Failure Modes}
\label{sec:wearable}
A wearable has to work in conditions that retrospective datasets rarely contain, and the move from cuff-based to contactless monitoring raises deployment questions beyond BP alone \cite{man2022}. Walking, exercise, sleep movement, sweating, temperature change, sensor repositioning, and inconsistent strap tightness all degrade PPG quality \cite{allen2007,sun2016,zhao2023}, and they corrupt the very landmarks BP estimation depends on. Even for heart rate, the simplest PPG measurement, error during activity was about 30\% higher than at rest across commercial devices \cite{bent2020}.

Motion matters most because continuous monitoring is most valuable when the wearer is moving. An algorithm that discards most data during activity or sleep transitions fails exactly when continuous information is needed. Contact pressure causes a related problem: loose contact increases motion artifact, while excessive pressure alters local perfusion and can mimic a BP change when central pressure has not moved. Posture has a measurable effect as well. In a validation of a commercial optical bracelet, agreement with auscultation met ISO 81060-2 criteria when participants sat with the wrist at heart level, but the standard deviation of the error rose to about 10\,mmHg lying down and about 12.5\,mmHg standing \cite{sola2021}. Cold exposure, dehydration, and peripheral vascular disease further reduce PPG amplitude independently of central BP, and wrist, finger, ear, and phone-based sensors couple to tissue differently, so results from one site should not be assumed to hold at another.

Signal-quality assessment is the first line of defense. A signal-quality index computed from waveform morphology, spectral content, or an accelerometer can flag or correct corrupted segments before they reach the BP estimator. Pollreisz and TaheriNejad compared detection and removal strategies, from morphological heuristics to reference-signal and accelerometer-based methods, and found that no single technique performs well across all activity types \cite{pollreisz2022}. Errors in BP estimation may have greater clinical consequences than errors in heart-rate estimation, because a beat that passes the quality filter with subtly distorted morphology can still yield a plausible but wrong BP value. Signal-quality assessment belongs in the reported estimation pipeline, together with the share of data it rejects.

\begin{table}[!t]
\caption{Real-World Conditions That Should Be Tested in Wearable Cuffless BP Studies}
\label{tab:conditions}
\centering
\footnotesize
\renewcommand{\arraystretch}{1.10}
\setlength{\tabcolsep}{3pt}

\begin{tabular}{|
>{\raggedright\arraybackslash}p{0.36\columnwidth}|
>{\raggedright\arraybackslash}p{0.56\columnwidth}|}
\hline
\textbf{Condition} & \textbf{Reason for inclusion} \\
\hline
Resting seated posture & Baseline comparison with conventional measurement \\
\hline
Standing posture & Tests hydrostatic and vascular changes \\
\hline
Walking or arm movement & Tests motion robustness \\
\hline
Exercise or recovery & Tests BP increase and dynamic tracking \\
\hline
Sleep or supine rest & Tests nocturnal monitoring feasibility \\
\hline
Cold or low-perfusion state & Tests weak PPG signal conditions \\
\hline
Repeated use over time & Tests drift and recalibration needs \\
\hline
\end{tabular}
\end{table}

A robust device has to be validated under its intended conditions of use, not only at rest. Table~\ref{tab:conditions} lists the conditions wearable studies should test. Resting seated measurement provides the baseline comparison with conventional devices; standing, walking or arm movement, exercise and recovery, sleep, and cold or low-perfusion states test robustness where PPG degrades; and repeated use over time tests drift and recalibration needs.

\section{Clinical Translation}
\label{sec:clinical}
\subsection{From Benchmark Accuracy to Clinical Meaning}
\begin{table*}[!t]
\caption{Evidence Pathway for Clinical Translation of Cuffless BP Technology}
\label{tab:pathway}
\centering
\footnotesize
\renewcommand{\arraystretch}{1.10}
\setlength{\tabcolsep}{3pt}

\begin{tabular}{|
>{\raggedright\arraybackslash}p{0.18\textwidth}|
>{\raggedright\arraybackslash}p{0.28\textwidth}|
>{\raggedright\arraybackslash}p{0.26\textwidth}|
>{\raggedright\arraybackslash}p{0.22\textwidth}|}
\hline
\textbf{Stage} & \textbf{Evidence needed} & \textbf{Main question} & \textbf{Claim supported} \\
\hline
Retrospective benchmark & Curated waveform data with clear splits & Can the algorithm estimate BP on existing data? & Feasibility \\
\hline
Controlled validation & Standardized posture, activity, and BP-change testing & Does the device work under controlled conditions? & Technical validation \\
\hline
Ambulatory validation & Daily-life testing with realistic motion and sensor use & Does the device work outside the lab? & Wearable-use validation \\
\hline
Longitudinal validation & Repeated testing over time after calibration & Does accuracy remain stable? & Drift and recalibration evidence \\
\hline
Clinical outcome evaluation & Diagnosis, treatment, adherence, or risk-prediction studies & Does the device improve care? & Clinical utility \\
\hline
\end{tabular}
\end{table*}

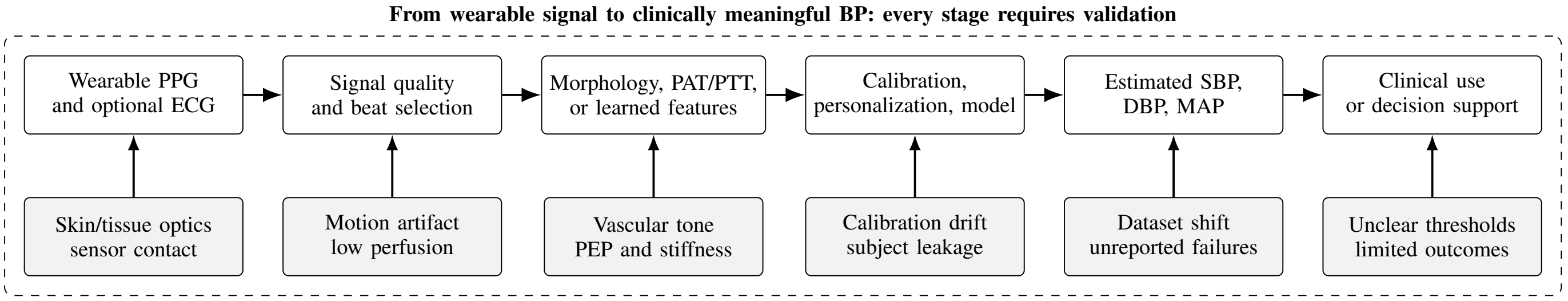
\begin{figure*}[t]
\centering
\resizebox{\textwidth}{!}{%
\begin{tikzpicture}[
  node distance=6mm and 5mm,
  every node/.style={font=\footnotesize},
  stage/.style={
    draw,
    rounded corners=2pt,
    align=center,
    minimum width=2.8cm,
    minimum height=10mm,
    inner sep=3pt,
    fill=white
  },
  risk/.style={
    draw,
    rounded corners=2pt,
    align=center,
    minimum width=2.8cm,
    minimum height=10mm,
    inner sep=3pt,
    fill=gray!10
  },
  arr/.style={-{Latex[length=2.5mm]}, thick},
  riskarr/.style={-{Latex[length=2mm]}, thick}
]

\node[stage] (n1) {Wearable PPG\\and optional ECG};
\node[stage, right=of n1] (n2) {Signal quality\\and beat selection};
\node[stage, right=of n2] (n3) {Morphology, PAT/PTT,\\or learned features};
\node[stage, right=of n3] (n4) {Calibration,\\personalization, model};
\node[stage, right=of n4] (n5) {Estimated SBP,\\DBP, MAP};
\node[stage, right=of n5] (n6) {Clinical use\\or decision support};

\node[risk, below=8mm of n1] (c1) {Skin/tissue optics\\sensor contact};
\node[risk, below=8mm of n2] (c2) {Motion artifact\\low perfusion};
\node[risk, below=8mm of n3] (c3) {Vascular tone\\PEP and stiffness};
\node[risk, below=8mm of n4] (c4) {Calibration drift\\subject leakage};
\node[risk, below=8mm of n5] (c5) {Dataset shift\\unreported failures};
\node[risk, below=8mm of n6] (c6) {Unclear thresholds\\limited outcomes};

\draw[arr] (n1) -- (n2);
\draw[arr] (n2) -- (n3);
\draw[arr] (n3) -- (n4);
\draw[arr] (n4) -- (n5);
\draw[arr] (n5) -- (n6);

\draw[riskarr] (c1) -- (n1);
\draw[riskarr] (c2) -- (n2);
\draw[riskarr] (c3) -- (n3);
\draw[riskarr] (c4) -- (n4);
\draw[riskarr] (c5) -- (n5);
\draw[riskarr] (c6) -- (n6);

\begin{scope}[on background layer]
\node[
  draw,
  dashed,
  rounded corners=3pt,
  fit=(n1)(n6)(c1)(c6),
  inner sep=7pt,
  label={[font=\footnotesize\bfseries]above:
  {From wearable signal to clinically meaningful BP: every stage requires validation}}
] {};
\end{scope}

\end{tikzpicture}
}
\caption{Graphical summary of the survey. PPG-based cuffless BP estimation is best understood as a measurement chain rather than a single regression task.}
\label{fig:summary}
\end{figure*}

Clinical usefulness depends on whether BP estimates remain valid as the user, sensor, physiology, environment, and calibration state change. Fig.~\ref{fig:summary} draws the preceding sections together. Its upper row follows the chain from a wearable PPG signal to clinical use, and its lower row names the main risk at each stage: skin optics and sensor contact, motion artifact and low perfusion, vascular tone and PEP, calibration drift and subject leakage, dataset shift and unreported failures, and unclear thresholds with limited outcome evidence. A weakness at any stage can make an accurate benchmark result clinically unreliable.

The evidence from commercial devices bears this out. A 2022 search of 205 commercial cuffless devices found that only 11 had undergone any validity testing and none had been tested under IEEE 1708 \cite{cox2022}. A meta-analysis of 16 validation studies covering 15 devices and 974 participants found pooled mean differences of 3.42\,mmHg for SBP and 1.16\,mmHg for DBP, but with very high heterogeneity ($I^2$ of 95\% and 87\%) and without assessing the spread of errors that ISO criteria require \cite{islam2022}. A small average bias, in other words, coexists with large and inconsistent errors across devices and studies.

Clinical decisions are also asymmetric in their risks. Underestimating high systolic BP may delay treatment, while overestimation causes unnecessary concern. A trend tracker, a home hypertension support tool, and a medication-titration system need different evidence. A trend tracker that reports relative changes for general wellness must show that its readings move in the right direction under everyday conditions. A home support tool that prompts users to confirm elevated readings with a cuff must detect elevated BP reliably and report how often it raises false alarms. A system used to adjust medication must agree with reference measurements in absolute mmHg across the hypertensive range, stay stable between calibrations, and be supported by outcome evidence. Continuous data may one day reveal nocturnal patterns, morning surge, and medication response, but the rules for interpreting them must be validated clinically, not borrowed from intermittent cuff thresholds.

\subsection{Evidence Required at Each Stage}
\label{sec:translation}
Good performance on a curated waveform dataset does not turn a model into a clinical measurement device. Table~\ref{tab:pathway} pairs each stage of the validation ladder in Fig.~\ref{fig:ladder} with the evidence it needs, the question it answers, and the claim it supports. A retrospective benchmark shows whether an algorithm can estimate BP on existing data and supports only a feasibility claim. Controlled and ambulatory validation show whether the device works under standardized conditions and in daily life, longitudinal validation whether its accuracy holds, and only clinical outcome studies whether it improves care. A 2025 international review in \textit{JAMA Cardiology} reached the same conclusion: cuffless devices are appealing but need validation before clinical deployment, with open concerns about accuracy, undertreatment or overtreatment, equitable access, privacy, and workflow integration \cite{yang2025}. The claim made for a device should match the stage of evidence actually behind it.

\section{Future Research Directions}
\label{sec:future}
Future progress will require stronger evidence in addition to further improvements in benchmark accuracy. Important gaps remain in longitudinal datasets, uncertainty estimation, and standardized reporting. The first is data. Aurora-BP shows that large ambulatory collections with cuff references can be built and shared \cite{mieloszyk2022}, but to our knowledge no public dataset yet follows participants over weeks to months with repeated reference measurements, several device types, and documented skin tone, which together are needed to study drift and bias. Such a benchmark should also ship with predefined subject-wise partitions so that results can be compared directly. The second is uncertainty. Models should report how confident they are in each estimate, for example as a prediction interval or a drift flag, so that readings taken under poor signal quality or after the calibration has degraded can be identified. Shen et al.\ have shown that conformal prediction can attach intervals with close to nominal coverage to ambulatory cuffless estimates \cite{shen2025}. The third is reporting. BP-range- and skin-tone-stratified error tables should be a condition of publication, much as CONSORT-style checklists are for clinical trials, and the pulse-oximetry experience shows why.

Several technical directions may help address these limitations. Physics-informed models that build the propagation and reflection mechanisms described earlier into learning have already reduced labeling needs in related signals \cite{sel2023} and could make failures easier to diagnose. Continual personalization, which updates a model as new weakly labeled data arrive from a user or sensor, offers a middle ground between calibration-free and fully individualized models. Multimodal sensing that fuses PPG with accelerometry, temperature, or bioimpedance on the device could address motion, contact pressure, and skin temperature at the source, at the cost of hardware complexity and power; on-device optimization of the kind reviewed earlier makes this increasingly practical.

\begin{table}[!t]
\caption{Minimum Reporting Checklist for Future Studies}
\label{tab:checklist}
\centering
\footnotesize
\renewcommand{\arraystretch}{1.25}
\setlength{\tabcolsep}{4pt}

\begin{tabular}{|
>{\raggedright\arraybackslash}p{0.29\columnwidth}|
>{\raggedright\arraybackslash}p{0.63\columnwidth}|}
\hline
\textbf{Item} & \textbf{Required detail} \\
\hline
Population & Age, sex, health status, BP range \\
\hline
Sensor site & Wrist, finger, ear, phone, or other \\
\hline
Signals & PPG only or multimodal inputs \\
\hline
Reference BP & Cuff, arterial line, or other source \\
\hline
Calibration & Initial, repeated, or calibration-free \\
\hline
Split design & Subject-independent and temporal policy \\
\hline
Motion handling & Inclusion, exclusion, and failure rate \\
\hline
Metrics & Mean error, SD, MAE, RMSE, stratified error \\
\hline
Deployment claim & Screening, tracking, or clinical measurement \\
\hline
\end{tabular}
\end{table}

Reporting should finally be complete enough for a reader to tell which tier of evidence a study reaches. Table~\ref{tab:checklist} lists the minimum information future studies should report: study population, sensor site, input signals, reference BP source, calibration approach, split design, motion handling and failure rate, error metrics, and the intended deployment claim, whether screening, tracking, or clinical measurement.

\section{Limitations of This Survey}
\label{sec:limitations}
This review has three main limitations. It is a critical narrative review, not a registered systematic review. Retrieval was targeted and not exhaustively logged, so we cannot report the numbers of records identified, screened, and excluded, as a PRISMA-style review would, and readers cannot confirm that no significant contrary evidence was missed. We also did not pool reported accuracy values, because differences in dataset, population, calibration protocol, and train--test design would make a pooled estimate misleading. As a result, the review cannot say whether one family of methods reliably outperforms another, only which comparisons are well or poorly supported. The review emphasizes measurement validity when interpreting benchmark accuracy. This reflects the analytical framework adopted in this survey rather than a universally accepted prioritization. Our claims rest on the cited evidence, but broader normative positions, such as the view that the field should invest in validation infrastructure before further model refinement, are our assessment and not an established consensus. A systematic review with a registered protocol and strict inclusion criteria, limited to studies with transparent subject-independent evaluation and reproducible calibration, BP-range, and failure-rate reporting, would complement this work.

\section{Conclusion}
\label{sec:conclusion}
PPG-based cuffless BP monitoring is attractive because it could provide frequent, comfortable, and potentially continuous information that intermittent cuff readings miss. It remains, however, an unsolved measurement problem. PPG and pulse-timing signals are indirect surrogates for arterial pressure, and a low reported error does not rule out the confounders reviewed here. Where models have been tested outside their training data, in other datasets, in ambulatory use, or as commercial devices, errors have been several times larger than the benchmark literature suggests. For researchers, this means designing cuffless BP studies as measurement studies before presenting them as prediction studies. For clinicians and device developers, it means treating a calibrated trend estimate as something other than a diagnostic BP measurement and matching every claim to the evidence behind it.

With roughly four in five of the 1.4 billion adults with hypertension lacking adequate control \cite{who2024}, a device deployed on the strength of a retrospective benchmark alone could delay diagnosis or misdirect treatment and widen the very gap it was meant to close. Clinical readiness depends not only on benchmark MAE, but also on whether the complete measurement chain remains reliable across users, sensors, time, and real-world conditions. Current evidence therefore supports PPG-based cuffless BP estimation as a promising research direction, while further validation is required before dependable clinical use can be established.

\end{document}